\documentclass[12pt]{article}
\usepackage{cite,epsfig,amssymb,amsmath,graphicx,color}
\usepackage{subfigure, float}
\usepackage{caption}
\usepackage{kotex}
	\DeclareCaptionFormat{myformat}{#3}
\newcommand{\nn}{\nonumber}

\newcommand{\1}{1\kern -3pt \mathrm{l}}

\begin{document}
        
        \vspace{5mm}
        \begin{center}
                {{\Large \bf Non-Relativistic Supersymmetric Gauge Theories and Soliton Solutions  }}
                \\[17mm]
                Yoonbai Kim, ~~O-Kab Kwon, ~~D. D. Tolla \\ [2mm]
                {\it Department of Physics, Sungkyunkwan University, Suwon 16419, Korea} \\ [-1mm]
                {\it yoonbai@skku.edu, ~okab@skku.edu, ~\it ddtolla@skku.edu} \\ [3mm]
                Hanwool Song\\ [2mm]
                {\it School of Physics, Korea Institute for Advanced Study, 85 Hoegiro, Dongdaemun-Gu, Seoul 02455, Korea} \\ [-1mm]
                {\it hanwoolsong0@gmail.com} \\ [3mm]
                
        \end{center}
        \vspace{15mm}

        \begin{abstract}
This paper investigates non-relativistic supersymmetric gauge theories obtained through null reduction, with particular focus on supersymmetric Abelian Higgs models and their soliton solutions. Starting from four-dimensional ${\cal N} = 1$ relativistic supersymmetric theories, we apply null reduction to derive Galilean-invariant non-relativistic theories in $2 + 1$ dimensions while preserving gauge symmetry and supersymmetry. The resulting symmetry structure is shown to correspond to the Bargmann extension of the super-Galilean algebra. We first analyze the massive supersymmetric Abelian Higgs model and show that, although it possesses nontrivial vacuum configurations, the mass term obstructs the existence of nontrivial BPS vortex solutions. We then study the massless supersymmetric Abelian Higgs model with multiple charged scalar fields, for which the BPS equations admit vortex-type soliton solutions under suitable charge and winding number conditions. The vacuum moduli space is identified as a weighted projective space, leading naturally to both integer and fractional vortex configurations. By imposing radial symmetry, we derive the corresponding vortex equations and examine their solutions analytically and numerically. Finally, we relate the vacuum structure of the model to the Ginzburg–Landau theory of superconductivity, highlighting how familiar superconducting behavior can emerge from a symmetry-preserving non-relativistic supersymmetric framework.     
        \end{abstract}

\newpage
  \tableofcontents

\section{Introduction}

Relativistic field theories are strongly constrained by symmetry principles such as gauge symmetries, global internal symmetries, spacetime symmetries, and supersymmetry. These symmetries determine the structure of admissible Lagrangians and often provide analytic control over non-perturbative objects such as solitons. In many physical situations, however, the relevant dynamics are effectively non-relativistic, as in condensed matter systems and low-energy limits of quantum field theories. It is therefore important to take non-relativistic limits in a way that preserves the essential symmetry structure of the relativistic parent theory. If such limits are taken in an uncontrolled manner, gauge invariance or supersymmetric constraints may be lost or obscured \cite{Jose:1991A,ClarkLove:1984,Meyer:2017zfg,Baiguera:2024}. By contrast, symmetry-preserving reductions allow the conceptual and technical advantages of relativistic theories to survive in the non-relativistic regime \cite{Bagchi:2022twx,Auzzi:2019kdd,Baiguera:2022cbp}.

A particularly useful framework for this purpose is null reduction. In this approach, a non-relativistic theory in $(d+1)$ dimensions is obtained from a relativistic theory in $(d+2)$ dimensions by compactifying along a null direction. The resulting theory is naturally Galilean invariant, with the null momentum becoming the central charge of the Bargmann algebra. Moreover, gauge symmetry and supersymmetry descend in a controlled way. For gauge theories, null reduction leads naturally to the magnetic non-relativistic limit, in which the gauge field strength plays the role of an external magnetic background, in close analogy with the non-relativistic limit of gravity and the magnetic limit of electrodynamics \cite{BellacLevyLeblond,Festuccia:2016caf,HollandBrown}.

In this paper, we construct the non-relativistic supersymmetric Abelian Higgs models in $(2+1)$ dimensions by applying null reduction to relativistic theories with supersymmetry. We then investigate
the possible vortex solutions supported by the resulting nonrelativistic
theories.
Supersymmetry is especially valuable in the study of soliton solutions, since
the vanishing of supersymmetry variations, or equivalently the Killing spinor
equations, naturally leads to Bogomol'nyi bounds and first-order
Bogomol'nyi--Prasad--Sommerfield (BPS) equations
\cite{BPS1,BPS2}. Vortex solutions in Abelian gauge theories have a long
history, beginning with Abrikosov vortices in type-II superconductors and their
field-theoretic realization as Nielsen--Olesen vortices in the Abelian Higgs
model \cite{Abrikosov:1956sx,Nielsen:1973cs}. The existence and uniqueness of
topological vortices were subsequently established by Taubes
\cite{Taubes:1979tm}.

 We first consider the massive supersymmetric Abelian Higgs model. Although this model admits nontrivial constant extrema, the mass term obstructs the existence of nontrivial BPS vortex configurations. We therefore focus on the massless model with two chiral fields carrying different gauge couplings. In this case, the BPS equations admit vortex-type solutions, and the vacuum structure plays a central role in classifying them.

When the two gauge couplings are parametrized in terms of integer charges as $g_a = g_0 q_a$ with $q_a \in \mathbb{Z}^{+}$, 
where \(g_0\) denotes the minimal unit of charge, the vacuum moduli space is
identified with the weighted projective line $W\mathbb{CP}^1_{(q_1,q_2)}$
~\cite{Eto:2009bz}. This geometric structure determines the allowed
asymptotic behavior of the scalar fields and, consequently, the possible
vortex configurations.
At a generic point of the vacuum moduli space, where both scalar fields have nonzero vacuum expectation values, finite energy condition requires their asymptotic phases to wind along the same broken U(1) gauge orbit. This requirement
imposes the compatibility condition $\frac{n_1}{g_1} = \frac{n_2}{g_2}$ 
or equivalently, $\frac{n_1}{q_1} = \frac{n_2}{q_2}$, 
where $n_1$ and $n_2$ are the winding numbers of the two scalar fields.
In this case, the corresponding vortex configurations carry ordinary
integer vortex number.
At the orbifold fixed points of \(W\mathbb{CP}^1_{(q_1,q_2)}\), however,
one of the scalar vacuum expectation values vanishes, and a residual discrete
gauge symmetry remains unbroken. As a result, the same BPS equations admit
solutions with fractional vortex number, giving rise to fractional vortex
configurations.

Multicomponent superconducting systems provide an important physical motivation for studying Abelian Higgs-type models with more than one charged scalar field. The possibility of superconductivity with multigap was discussed in \cite{Suhl:1959zz}, while fractional-flux vortices in two-gap superconductors were studied in \cite{Babaev:2001hv}.  From the high-energy field theory perspective, semilocal strings provide a closely related class of vortex solutions in Abelian gauge theories with multiple scalar fields \cite{Vachaspati:1991dz}. Although our model is not a microscopic multiband superconductor model, these works clarify the broader physical context of multiscalar Abelian Higgs systems and fractional vortex configurations.

To obtain explicit solutions, we impose radial symmetry on the BPS equations. For generic asymptotic vacua, the radial ansatz yields topological vortex solutions subject to the winding--charge compatibility condition. For asymptotic vacua located at the orbifold fixed points, the same equations admit fractional winding solutions. We analyze their asymptotic behavior and present numerical profiles for the scalar fields, magnetic field, and energy density. We also compare the vacuum structure of our non-relativistic supersymmetric Abelian Higgs models with that of the Ginzburg--Landau theory of superconductivity \cite{GLT}. Recent field-theoretic approaches to non-relativistic superconductivity and charged vortices have been developed in \cite{Jeon:2025snd,Kim:2025khe,Kim:2025mml,Kim:2025ien} for both BPS and non-BPS cases. Our construction differs from these works in that it is obtained by null reduction from a supersymmetric relativistic parent theory, while preserving both gauge symmetry and supersymmetry in a controlled non-relativistic framework.

The rest of this paper is organized as follows. In Section~2, we apply null reduction to the four-dimensional ${\cal N}=1$ super-Poincar\'e algebra and obtain the Bargmann extension of the super-Galilean algebra. In Section~3, we derive the non-relativistic supersymmetric Abelian Higgs models, verify their Galilean symmetry, and obtain the corresponding BPS equations. In Section~4, we classify and solve the vortex equations, emphasizing the role of the weighted projective vacuum moduli space and its orbifold fixed points in producing fractional vortices. Section~5 is devoted to conclusions. In the appendix, we discuss the relation between our vacuum solutions and those of the Ginzburg--Landau theory of superconductivity.

\section{Null Reduction and the non-Relativistic Limit}
Null reduction provides a systematic framework for deriving non-relativistic field theories in $(d+1)$ dimensions from relativistic theories in $(d+2)$ dimensions~\cite{Bagchi:2022twx,Auzzi:2019kdd,Baiguera:2022cbp}. Under this procedure, the relativistic symmetries are transmuted into Galilean symmetry. Moreover, when the parent $(d+2)$-dimensional relativistic theory 
enjoys conformal invariance, the null reduction induces 
Schr\"{o}dinger symmetry in the resulting $(d+1)$-dimensional 
non-relativistic theory. 
The Schr\"{o}dinger algebra thus obtained contains the Galilean 
algebra as a proper subalgebra.
In the supersymmetric context, the null reduction of the Super-Poincar\'{e} algebra yields the Bargmann extension of the Super-Galilean algebra.  Conceptually, the procedure realizes a non-relativistic limit by explicitly breaking Lorentz invariance and selecting a preferred temporal frame, and by transforming the relativistic mass-shell condition into a non-relativistic dispersion relation.
For a massless theory in $(d+2)$ dimensions, for instance, the relativistic mass-shell condition
\begin{align}
2 P_{+} P_{-} + \mathbf{P}_{\perp}^{2} = 0
\end{align}
reduces, upon fixing the null momentum $P_{-} \equiv M$, to the non-relativistic dispersion relation
\begin{align}
E = -P_{+} = \frac{\mathbf{P}_{\perp}^{2}}{2M}.
\end{align}
Here, $\mathbf{P}_{\perp}$ denotes the transverse spatial momentum, and 
$x^{\pm} = \frac{1}{\sqrt{2}}\left(x^{d+1} \pm x^{0}\right)$ are the light-cone coordinates. 
The null Killing vector $\xi = \partial_{-}$ generates translations along the compactified null direction, so that the corresponding momentum $P_{-}$ is identified with the non-relativistic mass operator $M$.
This reduction constrains the dynamics along the longitudinal light-cone direction, effectively mimicking low-velocity motion. Moreover, the mass parameter no longer simply labels states but emerges as a central charge in the Bargmann algebra, thereby ensuring the internal consistency of the non-relativistic symmetry structure.

More specifically, let us consider the null reduction of the ${\cal N}=1$ super-Poincar\'e algebra in four-dimensional space-time. By identifying the light-cone generator $P_{-}$ as a central charge, the reduction yields the non-relativistic super-Bargmann algebra in $(2+1)$ dimensions. We start from the ${\cal N}=1$ super-Poincar\'e algebra of the relativistic theory, which is given by
\begin{align}\label{SPalg}
	&[P_{m}, P_{n}] = 0,\qquad [M_{mn}, P_{r}] = -i\eta_{mr}P_{n}+i\eta_{nr}P_{m}, \nonumber\\
	&[M_{mn}, M_{rs}] = -i\eta_{mr}M_{ns}-i\eta_{ns}M_{mr}+i\eta_{ms}M_{nr}+i\eta_{nr}M_{ms}, \nonumber\\
	&\{Q_{\alpha},\bar Q_{\dot\alpha}\}=2\sigma^{m}_{\alpha\dot\alpha}P_{m},\qquad \{Q_{\alpha}, Q_{\beta}\}=0,\qquad [P_{m},Q_{\alpha}]=0, \nonumber\\	&[M_{mn},Q_{\alpha}]=i(\sigma_{mn})_{\alpha}{}^{\beta}Q_{\beta},
\end{align}
where $m,n,r,s=0,1,2,3$ label the space-time indices, $\alpha,\beta=1,2$ and $\dot\alpha,\dot\beta=\dot 1,\dot 2$ denote spinor indices, $\eta_{mn}=\mathrm{diag}(-,+,+,+)$ is the Minkowski metric, $P_{m}$ and $M_{mn}$ are the translation and Lorentz generators, and $Q_{\alpha}$ are the ${\cal N}=1$ supersymmetry generators.

In order to implement the null reduction for the 4-dimensional relativistic supersymmetric theory with the ${\cal N}=1$ super-Poincar\'e algebra given in \eqref{SPalg}, we introduce  light-cone coordinates $y^{m'}=(x^-,x^+,x^i)$,  $(i=1,2)$, defined by $x^{\pm}=\frac{1}{\sqrt2}(x^3\pm x^0)$. The corresponding light-cone momenta $P_{m'}=\frac{\partial x^m}{\partial y^{m'}}P_{m}=(P_-,P_+,P_i)$ with $P_{\pm}=\frac{1}{\sqrt2}(P_3\pm P_0)$. Similarly, the Lorentz generators transform as $M_{m'n'}=\frac{\partial x^m}{\partial y^{m'}}\frac{\partial x^n}{\partial y^{n'}}M_{mn}=(M_{-+},M_{-i},M_{+i},M_{ij})$,
where $M_{-+}=M_{30},~ M_{\pm i}=\frac{1}{\sqrt2}(M_{3i}\pm M_{0i})$.
The generators that commute with the null momentum $P_-$ are $\{ P_-, P_+, P_i, K_i \equiv M_{-i}, M_{ij} \}$, 
$i,j=1,2$, where in two spatial dimensions we define
$J \equiv M_{12}$.
These generators close into the $(2+1)$-dimensional Galilean algebra,
with $K_i$ identified as the Galilean boost generators and $J$ as the
spatial rotation generator.
The even part of the null-reduced Poincar\'e algebra,
namely the Bargmann algebra (the centrally extended Galilean algebra),
is then given by
\begin{align}\label{Baralg}
&[P_-, P_+] = 0, 
\qquad [P_-, P_i]=0, 
\qquad [P_-, K_i]=0, 
\qquad [P_-, J]=0, 
\nonumber \\[4pt]
&[P_+, P_i]=0, 
\qquad [P_+, K_i]= i P_i,
\qquad [P_+, J]=0, 
\nonumber\\[4pt]
&[P_i, P_j] = 0, 
\qquad [P_i, K_j] = - i \delta_{ij} P_-, 
\qquad [P_i, J]= i \epsilon_{ij} P_j, 
\nonumber \\[4pt]
&[K_i, K_j] =0, 
\qquad [K_i, J] = i \epsilon_{ij} K_j .
\end{align}
Here, we note  that the null momentum $P_-$ commutes with all generators, 
$[P_-,\, \mathcal{G}] = 0$, for any generator $\mathcal{G}$ of the algebra. Hence, $P_-$ plays the role of the central charge in the Bargmann extension of the Galilean algebra. The odd sector of the null-reduced super-Poincar\'e algebra is given by
\begin{align}\label{Qalgebra}
	&\{Q_1,\bar Q_{\dot1}\}=2\sqrt2 P_-,\quad \{Q_1,\bar Q_{\dot2}\}=2(P_1-iP_2),\quad \{Q_2,\bar Q_{\dot1}\}=2(P_1+iP_2),\quad \{Q_2,\bar Q_{\dot2}\}=-2\sqrt2 P_+\nn\\
	&[K_i,Q_1]=0,\quad [K_1-iK_2,Q_2]=i\sqrt2 Q_1,\quad [J,Q_1]=\frac 12 Q_1,\quad [J,Q_2]=-\frac 12 Q_2.
\end{align}

The above commutation and anticommutation relations demonstrate that
the resulting non-relativistic supersymmetry algebra is precisely the
supersymmetric extension of the Bargmann algebra, namely the centrally
extended Galilean algebra underlying non-relativistic physics. 
In this construction, we augmented the Bargmann generators, which are, Galilean boosts, spatial rotations, time and space translations, and the central mass generator, with the complex supercharge $Q_\alpha$. 
These supercharges transform as spinors under spatial rotations and obey anticommutation relations that close onto the Hamiltonian ($H=-P_+$) and, crucially, on the central charge $(P_-)$, reflecting the non-relativistic dispersion relation and the structure of Galilean invariance. Since the null reduction procedure preserves the number of fermionic
generators, the four-dimensional ${\cal N}=1$ super-Poincar\'e algebra,
containing four real supercharges, reduces to the non-relativistic
${\cal N}=2$ superalgebra in $(2+1)$ dimensions.

\section{Non-Relativistic Supersymmetric Gauge Theories}
The purpose of this section is to derive non-relativistic supersymmetric gauge theories via the null reduction of relativistic models, as introduced in the previous section.
We begin with a four-dimensional ${\cal N}=1$ supersymmetric Abelian Higgs model constructed in \cite{Kim:2023abp}.
Applying the null reduction procedure to this model, we obtain a non-relativistic supersymmetric Abelian Higgs model in $(2+1)$ dimensions.
By analyzing the resulting supersymmetry and spacetime transformation rules, we further show that the non-relativistic theory is invariant under the Bargmann extension of the Galilean transformations.

\subsection{Null-reduction of supersymmetric Abelian Higgs model}
Let us start by summarizing the standard relativistic $ {\cal N}=1$ supersymmetric Abelian Higgs model using the component field formalism. The field content of this model consists of a massless vector multiplet $(A_\mu,\lambda)$ and a pair of oppositely charged massive chiral multiplets $(\phi_a,\psi_a,~ a=1,2)$. 
The gaugino field $\lambda$ is a two component Majorana spinor, whereas the matter fermionic fields $\psi_a$ are two component Weyl spinors.  The Lagrangian of the  model is \cite{Kim:2023abp} 
\begin{align}\label{L-massive}
	{\cal L}=&-\frac1{4} F_{mn}F^{mn}-\frac i2\bar\lambda\bar\sigma^m \partial_m\lambda-D_m\bar\phi_a D^m\phi_a-i\bar\psi_a\bar\sigma^m D_m\psi_a-m^2\big|\phi_a|^2
	\nonumber\\
	&+\frac i2m_{ab}(\psi_a\psi_b-\bar\psi_a\bar\psi_b)+ig(-1)^a\big(\lambda\psi_a\bar\phi_a-\bar\lambda\bar\psi_a\phi_a\big)
	-\frac{g^2}2 \Big(|\phi_1|^2-|\phi_2|^2-\frac{\xi}g\Big)^2,
\end{align}
where, the $2\times 2$ mass matrix is $m_{ab} = \begin{pmatrix}
	0 & m\\
	m &0
\end{pmatrix}$ and the covariant derivatives $D_m$ for the two oppositely charged complex scalar fields and the two oppositely charged spinors are given by \begin{align}
	D_m \begin{pmatrix}
		\phi_a \\ \psi_a \end{pmatrix} = \big[\partial_{m} -(-1)^a ig A_m \big] \begin{pmatrix} \phi_a \\ \psi_a \end{pmatrix}.
\end{align}
The sigma-matrices, $\sigma^m$ and $\bar\sigma^m$,  are composed of the Pauli matrices $\sigma^i$ and the identity matrix ${\rm I}$ as $\sigma^m=(-{\rm I},\sigma^i),~\bar\sigma^m=(-{\rm I},-\sigma^i)$.
The Lagrangian  
\eqref{L-massive} is invariant under the ${\cal N}=1$ supersymmetry transformation:
\begin{align}\label{delta-3}
	&\delta_\epsilon\phi_a=\epsilon\psi_a, \qquad\delta_\epsilon\psi_a= i\sigma^m\bar\epsilon D_m\phi_a -im_{ab} \epsilon\bar\phi_b, \nn\\
	& \delta_\epsilon\lambda=-\sigma^{mn}\epsilon F_{mn}+ig\Big(|\phi_1|^2-|\phi_2|^2-\frac{\xi}g\Big)\epsilon,\qquad \delta_\epsilon A_m=-\frac {i}2(\bar\epsilon\bar\sigma_m\lambda-\bar\lambda\bar\sigma_m\epsilon).
\end{align} 
In addition, the Lagrangian is also invariant under space-time translations $x^m\to x^m+s^m$ and the Lorentz transformation $x^m\to x^m+\omega^m_{~n}x^n$, with the variations of the fields given by
\begin{align}
	&\delta_s\phi_a=-s^m\partial_m\phi_a, \qquad \delta_s\psi_a=-s^m\partial_m\psi_a,\quad \delta_s A_m=-s^n\partial_n A_m, \quad \delta_s\lambda=-s^m\partial_m\lambda, \label{sttr}\\
	&\delta_\omega\phi_a=-\omega^m_{~n}x^n\partial_m\phi_a, \quad \delta_\omega\psi_a=-\omega^m_{~n}x^n\partial_m\psi_a-\frac12\omega_{mn}\sigma^{mn}\psi_a,\quad\delta_\omega A_m=-\omega^n_{~l}x^l\partial_n A_m+\omega_{m}^{~n}A_n, \nn\\
	&\delta_\omega\lambda=-\omega^m_{~n}x^n\partial_m\lambda-\frac12\omega_{mn}\sigma^{mn}\lambda. \label{Lztr}
\end{align}
Under these transformations, the Lagrangian is invariant up to a total derivative term
\begin{align}\label{dL-maseless}
	(\delta_s+\delta_\omega){\cal L}=-\partial_m\big(s^m{\cal L}+\omega^m_{~n}x^n{\cal L}\big).
\end{align}

The null-reduction can be applied to obtain a $(d+1)$-dimensional Lagrangian,  whenever a $(d+2)$-dimensional theory has a $U(1)$ gauge symmetry. One can then propose an ansatz in which the $(d+2)$-dimensional fields are expressed as a special $U(1)$ transformations of the $d+1$-dimensional fields, where the parameter of the $U(1)$ transformations depend on the extra coordinate.  The transformation acts on the fields in the matter multiplets as
\begin{align}
	&\phi_1(x^-,x^\mu)=e^{iM x^-}\varphi_1(x^\mu),\quad \phi_2(x^-,x^\mu)=e^{-iM x^-}\varphi_2(x^\mu),\quad \psi_1(x^-,x^\mu)=e^{iM x^-}\chi_1(x^\mu),\nn\\
	& \psi_2(x^-,x^\mu)=e^{-iM x^-}\chi_2(x^\mu), 
\end{align}
where $M$ is the eigenvalue of the $U(1)$ mass operator $P_-$, and we have denoted the $(2+1)$-dimensional spacetime coordinates as $x^\mu=(t\equiv x^+,x^i)$. The $U(1)$ transformation  does not affect the fields in the gauge multiplet, so that 
\begin{align}
	\lambda\to\lambda,~A_-\to A_-\equiv N,~A_\mu\to A_\mu,\quad {\rm are ~independent~ of}~x^-.
\end{align}  
The null reduction of the Lagrangian is carried out by substituting these transformed fields into \eqref{L-massive}. Then, we obtain
\begin{align}\label{Nullm}
	{\cal L}_{nr}&=\frac12(\partial_t N)^2+\eta^{ij}E_i\partial_j N-\frac14 F_{ij}F^{ij}-\frac i2\bar\lambda\bar\sigma_- \partial_t\lambda-\frac i2\bar\lambda\bar\sigma^i \partial_i\lambda 
	\nonumber\\
& -i(-1)^a(M+gN)(\bar\varphi_a D_t\varphi_a
	-\bar D_t\bar\varphi_a \varphi_a)-\bar D_i\bar\varphi_a D^i\varphi_a-(-1)^a(M+gN)\bar\chi_a\bar\sigma_+\chi_a 
	\nonumber\\
& -i\bar\chi_a\bar\sigma_- D_t\chi_a-i\bar\chi_a\bar\sigma^i D_i\chi_a
	+\frac i2m_{ab}(\chi_a\chi_b-\bar\chi_a\bar\chi_b)+ig(-1)^a\big(\lambda\chi_a\bar\varphi_a-\bar\lambda\bar\chi_a\varphi_a\big) 
	\nonumber\\
& -m^2\big|\varphi_a|^2-\frac{g^2}2 \Big(|\varphi_1|^2-|\varphi_2|^2-\frac{\xi}g\Big)^2,
\end{align}
where we have replaced $E_i= F_{ti}$.  We note that the components $\lambda_2$ and $\chi_{a2}$ are not truly dynamical and can be integrated out.
Similarly, applying the null-reduction to the supersymmetry transformations, we obtain
\begin{align}\label{delta-4}
	&\delta_\epsilon\varphi_a=\epsilon\chi_a, \qquad  \delta_\epsilon\chi_a= (-1)^a\big(M+gN\big)\sigma_+\bar\epsilon\varphi_a+i\sigma_-\bar\epsilon D_t\varphi_a+i\sigma^i\bar\epsilon D_i\varphi_a -im_{ab} \epsilon\bar\varphi_b,\nn\\
	&\delta_\epsilon\lambda=-2\sigma_{-+}\epsilon\partial_t N+2\eta^{ij}\sigma_{+i}\epsilon\partial_j N-2\eta^{ij}\sigma_{-i}\epsilon E_j-\sigma^{ij}\epsilon F_{ij}+ig\Big(|\varphi_1|^2-|\varphi_2|^2-\frac{\xi}g\Big)\epsilon,\nn\\
	&\delta_\epsilon A_\mu=-\frac {i}2(\bar\epsilon\bar\sigma_\mu\lambda-\bar\lambda\bar\sigma_\mu\epsilon),\qquad \delta_\epsilon N=-\frac {i}2(\bar\epsilon\bar\sigma_-\lambda-\bar\lambda\bar\sigma_-\epsilon).
\end{align}

The bosonic symmetry transformations of the non-relativistic theory are obtained by performing the null reduction of the relativistic bosonic symmetries. 
In particular, the null reduction of space-time translations \eqref{sttr} leads to
\begin{align}\label{NRTrans}
	&\delta_s\varphi_a=-is_+ M\,\varphi_a - s_- \partial_t\varphi_a - s^i \partial_i\varphi_a, 
	\qquad 
	\delta_s\chi_a=-is_+ M\,\chi_a - s_- \partial_t\chi_a - s^i \partial_i\chi_a, \nn\\
	&\delta_s A_\mu=-s_- \partial_t A_\mu - s^i \partial_i A_\mu, 
	\qquad 
	\delta_s N=-s_- \partial_t N - s^i \partial_i N, 
	\qquad 
	\delta_s\lambda=-s_- \partial_t\lambda - s^i \partial_i\lambda .
\end{align}
Here, the parameter $s_+$ generates the central $U(1)$ transformation associated with the mass operator $M$ (in the massive model, for $\varphi_2$ and $\chi_2$, whose $P_-$ eigenvalue is $-M$, $M$ is replaced by $-M$ here and in \eqref{NRLrnz} below for overall sign convention), while $s_-$ and $s^i$ correspond to time and spatial translations in $(2+1)$ dimensions, respectively. 
Thus, the reduced transformations combine the global $U(1)$ symmetry with Galilean space-time translations.
For the Lorentz sector, the null reduction must be carried out in a manner compatible with the structure of the reduced symmetry algebra. 
In particular, the generators $M_{-+}$ and $M_{+i}$ do not survive in the null-reduced Lorentz algebra, which becomes the Bargmann extension of the Galilean algebra. 
Accordingly, all terms proportional to the parameters $\omega^{-+}$ and $\omega^{+i}$ are consistently omitted.
Applying this prescription to \eqref{Lztr}, we obtain
\begin{align}\label{NRLrnz}
	&\delta_\omega\varphi_a
	= i M \omega^{i-} x_i \varphi_a
	- \omega^{i-} t \partial_i \varphi_a
	- \omega^{i}{}_{j} x^j \partial_i \varphi_a, \nn\\
	&\delta_\omega\chi_a
	= i M \omega^{i-} x_i \chi_a
	- \omega^{i-} t \partial_i \chi_a
	- \omega^{i}{}_{j} x^j \partial_i \chi_a
	- \omega^{i-} \sigma^{i+} \chi_a
	- \frac{1}{2} \omega_{ij} \sigma^{ij} \chi_a, \nn\\
	&\delta_\omega A_i
	= - \omega^{j-} t \partial_j A_i
	- \omega^{j}{}_{k} x^k \partial_j A_i
	+ \omega^{i-} N
	+ \omega_{i}{}^{j} A_j, \nn\\
	&\delta_\omega N
	= - \omega^{i-} t \partial_i N
	- \omega^{i}{}_{j} x^j \partial_i N, 
	\qquad
	\delta_\omega A_t
	= - \omega^{i-} t \partial_i A_t
	- \omega^{i}{}_{j} x^j \partial_i A_t
	+ \omega^{-i} A_i, \nn\\
	&\delta_\omega \lambda
	= - \omega^{i-} t \partial_i \lambda
	- \omega^{i}{}_{j} x^j \partial_i \lambda
	- \omega^{i-} \sigma^{i+} \lambda
	- \frac{1}{2} \omega_{ij} \sigma^{ij} \lambda .
\end{align}

The null-reduced Lagrangian in \eqref{Nullm} is invariant under the transformations derived above. 
These coincide with the well-known Galilean transformations---more precisely, the Bargmann extension of the Galilean algebra---acting on scalar, fermionic, and vector fields. 
They preserve the Schr\"odinger equation for the scalar sector, the Pauli equation for the fermionic sector, and the magnetic-limit form of Maxwell's equations for the gauge field.
For instance, consider the equation of motion for the scalar field obtained from the null-reduced Lagrangian \eqref{Nullm}. 
The corresponding Schr\"odinger-type equation is

\begin{align}
	&2i(M+gN)D_t\varphi_a
	+ig(\partial_t N)\varphi_a
	-(-1)^a D_i D^i\varphi_a
	-ig\lambda\chi_a
	+(-1)^a m^2\varphi_a
	-g^2 \Big(|\varphi_1|^2-|\varphi_2|^2-\frac{\xi}{g}\Big)\varphi_a
	\nonumber \\
	&=0 .
\end{align}
This equation is manifestly invariant under the Bargmann-extended Galilean transformations given above.

The conjugate momenta for the dynamical fields are 
\begin{align}
	&\pi_{\varphi_a}=\frac{\partial {\cal L}_{nr}}{\partial\dot\varphi_a}=-i(-1)^a (M+gN)\bar\varphi_a,\quad \pi_{\bar\varphi_a}=\frac{\partial {\cal L}_{nr}}{\partial\dot{\bar\varphi}_a}=i(-1)^a (M+gN)\varphi_a, \nn\\ &\pi_{\chi_{a1}}=\frac{\partial {\cal L}_{nr}}{\partial\dot\chi_{a1}}=i\sqrt2\bar\chi_{a\dot1},\quad \pi_{\bar\chi_{a\dot1}}=\frac{\partial {\cal L}_{nr}}{\partial\dot{\bar\chi}_{a\dot1}}=0,\quad \pi_{\lambda_{1}}=\frac{\partial {\cal L}_{nr}}{\partial\dot\lambda_{1}}=\frac i{\sqrt2}\bar\lambda_{\dot1},\nn\\
	&\pi_{\bar\lambda_{\dot1}}=\frac{\partial {\cal L}_{nr}}{\partial\dot{\bar\lambda}_{\dot1}}= 0,\quad \pi_{A_i}=\frac{\partial {\cal L}_{nr}}{\partial\dot A_{i}}=\partial_i N,\quad\pi_{A_t}=\frac{\partial {\cal L}_{nr}}{\partial\dot A_{t}}=0,\quad \pi_{N}=\frac{\partial {\cal L}_{nr}}{\partial\dot N}=\partial_t N.
\end{align}
Then, the corresponding Hamiltonian density is given by 
\begin{align}\label{HamDen}
	{\cal H}&=\pi_{\varphi_a}\partial_t{\varphi}_a+ \pi_{\bar\varphi_a}\partial_t{\bar\varphi}_a+\pi_{\chi_{a1}}\partial_t{\chi}_{a1}+\partial_t{\bar\chi}_{a\dot1}\pi_{\bar\chi_{a\dot1}}+\pi_{\lambda_{1}}\partial_t\lambda_{1}+\partial_t\bar\lambda_{\dot1}\pi_{\bar\lambda_{\dot1}}+\pi_{A_i}\partial_t{A_i}+\pi_{N}\partial_t{N}-{\cal L}_{nr}\nn\\
	&=2g(M+gN)|\varphi_a|^2 A_t+\bar D_i\bar\varphi_a D^i\varphi_a+(-1)^a\sqrt2(M+gN)\bar\chi_{a\dot 2}\chi_{a2}-(-1)^a\sqrt2g\bar\chi_{a\dot 1}\chi_{a1}A_t\nn\\
	&-i\bar\chi_{a\dot 1}(D_1-iD_2)\chi_{a2}-i\bar\chi_{a\dot 2}(D_1+iD_2)\chi_{a1}-\frac i2\bar\lambda_{\dot 1}(\partial_1-i\partial_2)\lambda_2-\frac i2\bar\lambda_{\dot 2}(\partial_1+i\partial_2)\lambda_1\nn\\
	&+\frac12(\partial_t N)^2-\nabla^2 N A_t+\frac14 F_{ij}F^{ij}-\frac i2m_{ab}(\chi_a\chi_b-\bar\chi_a\bar\chi_b)-ig(-1)^a\big(\lambda\chi_a\bar\varphi_a-\bar\lambda\bar\chi_a\varphi_a\big)\nn\\
	&+m^2\big|\varphi_a|^2+\frac{g^2}2 \Big(|\varphi_1|^2-|\varphi_2|^2-\frac{\xi}g\Big)^2.
\end{align}

\subsubsection{Constant extremum points}
When the Fayet--Iliopoulos (FI) $D$-term is introduced into the massive Abelian Higgs model, supersymmetry is spontaneously broken. Then, there are no supersymmetric vacuum solutions with zero energy. Consequently, the vacuum structure must be determined by extremizing the Hamiltonian density, allowing for positive vacuum energy. To this end, we consider the bosonic sector of the Hamiltonian density in \eqref{HamDen}, setting all fermionic fields to zero. The Hamiltonian density then reduces to
\begin{align}
{\cal H}
&= 2g(M+gN)|\varphi_a|^2 A_t
  + \bar D_i \bar\varphi_a D^i \varphi_a
  + \frac12 (\partial_t N)^2
  - (\nabla^2 N) A_t
  + \frac14 F_{ij} F^{ij}
\nonumber\\
&\quad
  + m^2 |\varphi_a|^2
  + \frac{g^2}{2}
    \left(
        |\varphi_1|^2 - |\varphi_2|^2 - \frac{\xi}{g}
    \right)^2.
\end{align}
The temporal component of the gauge field, $A_t$, appears as a Lagrange multiplier in the Hamiltonian formulation. 
Its equation of motion therefore yields a constraint rather than a dynamical evolution equation. 
Varying the Hamiltonian density with respect to $A_t$ leads to
\begin{align}
2g(M+gN)\left(|\varphi_1|^2+|\varphi_2|^2\right)
- \nabla^2 N
= 0 .
\end{align}
Then the  Hamiltonian density reduces to
\begin{align}\label{HamDen-2}
	{\cal H}= |D_i\varphi_1|^2+|D_i\varphi_2|^2+\frac12(\partial_t N)^2+ m^2\big(|\varphi_1|^2+|\varphi_2|^2\big)+\frac{g^2}2\Big(|\varphi_1|^2-|\varphi_2|^2-\frac{\xi}g\Big)^2+\frac12B^2,
\end{align}
where $B=F_{12}$. 
Varying this Hamiltonian density with respect to each field, we obtain the following equations of extremum points, 
\begin{align}\label{EOMm}
	&\delta_{\varphi_1}{\cal H}=\left[-\bar D_i^2+m^2+g^2\Big(|\varphi_1|^2-|\varphi_2|^2-\frac{\xi}g\Big)\right]\bar\varphi_1=0,\nn\\
	&\delta_{\varphi_2}{\cal H}=\left[-\bar D_i^2+m^2-g^2\Big(|\varphi_1|^2-|\varphi_2|^2-\frac{\xi}g\Big)\right]\bar\varphi_2=0,\nn\\
	&\frac12\delta_{A_i}{\cal H}=g^2\big(|\varphi_1|^2+|\varphi_2|^2\big){\bf A}+g\big(|\varphi_1|^2{\bf\nabla}\Omega_1-|\varphi_2|^2{\bf\nabla}\Omega_2\big)+\frac12\nabla\times{\bf B}=0,\nn\\
	&\delta_{N}{\cal H}=-\partial_t^2N=0,
\end{align}
where in the third line, we have used $\varphi_a=|\varphi_a|e^{i\Omega_a}$. Constant solutions of these equations can be local minima or local maxima of the potential. In this section, we will sort them out by restricting the parameters of the model to the appropriate ranges.  

One of the possible solutions to the equations of extremum points in \eqref{EOMm} is the trivial solution  $\varphi_1=\varphi_2=0\Rightarrow$ $\nabla\times{\bf B}=0\Rightarrow {\bf B}=constant$ up to a pure gauge term. This gives the energy density of the trivial solution
\begin{align}\label{symvac}
	{\cal H}_{0}=\frac{\xi^2}2+\frac12H^2>0,
\end{align}
where we have set ${\bf B}=constant={\bf H}$, which is a uniform magnetic field.

Depending on whether $\xi$ is positive or negative, we can write two non-trivial constant solutions of those equations of extremum.   
Note that the third line of \eqref{EOMm} is the Ampere-Maxwell's equation of the gauge field, which we may rewrite as
\begin{align}
\frac12\nabla\times{\bf B}={\bf j},\quad {\rm with}~{\bf j}=-g^2\big(|\varphi_1|^2+|\varphi_2|^2\big){\bf A}-g\big(|\varphi_1|^2{\bf\nabla}\Omega_1-|\varphi_2|^2{\bf\nabla}\Omega_2\big).
\end{align}
For constant $|\varphi_1|$ and $|\varphi_2|$, this equation leads to the London equation (see below), and the solution of the London equation in the bulk of the space (far from the boundary) is a  vanishing magnetic field ${\bf B}=0$ and hence a vanishing vector potential ${\bf A}=0$ up to a pure gauge term. Then if we choose a positive $\xi$ such that  $\xi>\frac{m^2}{g}$, we can write the constant non-trivial solution of \eqref{EOMm} as,
\begin{align}
	|\varphi_1|^2=\frac{\xi}g-\frac{m^2}{g^2},\qquad \varphi_2=0.
\end{align} 
The energy density of this non-trivial state is given by
\begin{align}\label{nontriv+}
	{\cal H}_{vac}^+=\frac{m^2}g\left(\xi-\frac{m^2}{2g}\right),
\end{align}
which is also positive since $\xi>\frac{m^2}{g}$.

We note that if $0 \le \xi \le \frac{m^2}{g}$, the only solution is the trivial one. 
On the other hand, if $\xi > \frac{m^2}{g}$, both the trivial solution and the non-trivial solutions exist as stationary solutions of the equations of motion. 
To compare their energies, let us parametrize $\xi = (1-\alpha)\frac{m^2}{g}$, where $\alpha$ is a negative real number $(\alpha<0)$. 
Then the energy density of the non-trivial solution \eqref{nontriv+} and that of the trivial solution \eqref{symvac} can be written as
\begin{align}
	{\cal H}_{vac}^+=(1-2\alpha)\frac{m^4}{2g^2}, \qquad {\cal H}_{0}=\left(1-2\alpha+\alpha^2\right)\frac{m^4}{2g^2}+\frac12H^2={\cal H}_{vac}^++\frac{\alpha^2m^4}{2g^2}+\frac12H^2. 
\end{align}
Therefore, the non-trivial solution has lower energy than the trivial one and is energetically preferred.

Next, we duplicate the above results for $\xi<0$. To do that, we assume $\xi<-\frac{m^2}{g}$, so that we can write another non-trivial vacuum solution,
\begin{align}
	\varphi_1=0, \qquad |\varphi_2|^2=-\frac{m^2}{g^2}-\frac{\xi}g,
\end{align} 
where again the magnetic field ${\bf B}$ and the vector potential ${\bf A}$ are set to zero by using the Ampere-Maxwell's equation.
The energy of this non-trivial solution is also positive and given by
\begin{align}
	{\cal H}_{vac}^-=-\frac{m^2}g\left(\xi+\frac{m^2}{2g}\right), 
\end{align}
which is positive since $\xi<-\frac{m^2}{g}$.  In this case as well, the only possible solution is the trivial one, if $~-\frac{m^2}{g}\le\xi\le 0$. However, if $\xi<-\frac{m^2}{g}$, both the trivial solution and non-trivial solution are possible solutions.  In order to show that the non-trivial solution has the lower energy, we can set $\xi=-(1-\alpha)\frac{m^2}{g}$, for $(\alpha<0)$, and express the energies as 
\begin{align}
	{\cal H}_{vac}^-=(1-2\alpha)\frac{m^4}{2g^2},\qquad {\cal H}_{0}=\left(1-2\alpha+\alpha^2\right)\frac{m^4}{2g^2}+\frac12H^2={\cal H}_{vac}^-+\frac{\alpha^2m^4}{2g^2}+\frac12H^2.
\end{align}
Again, as compared to the trivial solution, the  non-trivial solution is the favourable lower energy stable vacuum state.

\subsubsection{London equations}
In order to derive the Ampère–Maxwell equation for the gauge field, 
let us first expand the Hamiltonian density \eqref{HamDen-2} as
\begin{align}
    {\cal H} 
    &= (D_i\varphi_1)^\dagger (D_i\varphi_1)
      + (D_i\varphi_2)^\dagger (D_i\varphi_2)
      + \frac12 (\partial_t N)^2
      + \frac14 F_{ij}F^{ij}
\nonumber\\
    &\quad
      + m^2 \left(|\varphi_1|^2 + |\varphi_2|^2\right)
      + \frac{g^2}{2}
      \left(|\varphi_1|^2 - |\varphi_2|^2 - \frac{\xi}{g}\right)^2 .
\end{align}
Then, we parametrize the scalar fields as
\begin{align}
    \varphi_a = |\varphi_a| e^{i\Omega_a}, 
    \qquad a=1,2 .
\end{align}
Extremizing the Hamiltonian with respect to $A_i$ yields
\begin{align}
    \frac12\frac{\delta {\cal H}}{\delta A_i}
    = g^2 \big(|\varphi_1|^2 + |\varphi_2|^2\big) A_i
      + g \big(|\varphi_1|^2 \partial_i \Omega_1
      - |\varphi_2|^2 \partial_i \Omega_2\big)
      + \frac{1}{2}\partial_j F_{ij}
    = 0 .
\end{align}
In vector notation, this becomes
\begin{align}
    g^2 \big(|\varphi_1|^2 + |\varphi_2|^2\big)\mathbf{A}
    + g \big(|\varphi_1|^2 \nabla \Omega_1
    - |\varphi_2|^2 \nabla \Omega_2\big)
    + \frac{1}{2} \nabla \times \mathbf{B}
    = 0 .
\end{align}
This is the Ampère–Maxwell's law 
\begin{align}
	\frac12\nabla\times{\bf B}={\bf j}, \qquad {\rm with}~{\bf j}=-g^2\big(|\varphi_1|^2+|\varphi_2|^2\big){\bf A}-g\big(|\varphi_1|^2{\bf\nabla}\Omega_1-|\varphi_2|^2{\bf\nabla}\Omega_2\big).
\end{align}
Inserting the constant vacuum solutions of the previous subsection, we can write 
\begin{align}
	&g^2|\varphi_1|^2\nabla\times{\bf A}+\frac12\nabla\times(\nabla\times{\bf B})=0\Rightarrow g^2|\varphi_1|^2{\bf B}-\frac12\nabla^2{\bf B}=0, \quad {\rm when}~\xi>0\nn\\
	&g^2|\varphi_2|^2\nabla\times{\bf A}+\frac12\nabla\times(\nabla\times{\bf B})=0\Rightarrow g^2|\varphi_2|^2{\bf B}-\frac12\nabla^2{\bf B}=0, \quad {\rm when}~\xi<0.
\end{align}
Both equations take the form of the London equation. 
Identifying the coefficient of $\mathbf{B}$ with the inverse squared 
London penetration depth, we obtain
\begin{align}
    \lambda_L^{-2}
    = 2 g^2 |\varphi_a|^2
    = - 2 \alpha m^2
    = 2 |\alpha| m^2 ,
\end{align}
where we have used $\alpha<0$ in the broken phase.
Accordingly, the magnetic field satisfies
\begin{align}
    \nabla^2 \mathbf{B} = \lambda_L^{-2}\mathbf{B},
\end{align}
whose solution is exponentially damped,
\begin{align}
    \mathbf{B}(x) \sim \mathbf{B}_0 \, e^{-x/\lambda_L}.
\end{align}

\subsubsection{BPS equations}\label{BPSEQ}
Finding the non-constant solutions to the second order differential equations in \eqref{EOMm} is quite challenging. Since static (time independent) solutions of the corresponding BPS equations are also solutions to the equations in \eqref{EOMm}, one can instead solve those BPS equations to overcome this challenge. The standard procedure to derive the BPS equations is by reshuffling the bosonic part of the Hamiltonian and cast it as the sum of complete square terms and a boundary term.

In general, the BPS equations are obtained by saturating the energy bound, that is, by minimizing the Hamiltonian and requiring each complete-square term 
to vanish. In the present model, however, supersymmetry provides a more direct and systematic approach. Instead of relying solely on the Bogomol'nyi completion, we derive the BPS equations from the Killing spinor equations, which correspond 
to the vanishing of the fermionic supersymmetry variations.
Since the half BPS states preserve only half of the supersymmetry, we proceed by choosing the supersymmetry parameter $\epsilon$, as follows:
\begin{align}\label{ealp}
\epsilon_\alpha=\begin{pmatrix}
\chi\\
0
\end{pmatrix}\quad\Longleftrightarrow
\quad\epsilon^\alpha= \varepsilon^{\alpha \beta} \epsilon_\beta  =\begin{pmatrix}
0\\
\chi
\end{pmatrix},
\qquad\bar\epsilon_{\dot\alpha}=\begin{pmatrix}
\bar\chi\\
0
\end{pmatrix},\qquad\bar\epsilon^{\dot\alpha}= \varepsilon^{\dot \alpha \dot \beta}\bar \epsilon_{\dot \beta} =\begin{pmatrix}
0\\
\bar\chi
\end{pmatrix},
\end{align}
where $\varepsilon^{12} = \varepsilon^{\dot 1\dot 2}=-1$ and $\chi$, $\bar \chi$ are complex Grassmann parameters. Then, the sigma matrices act on $\epsilon$ as follows
\begin{align}
\sigma_+\bar\epsilon=0, \qquad \sigma_-\bar\epsilon=-\sqrt 2\,\bar\epsilon,\qquad \sigma^2\bar\epsilon=-i\sigma^1\bar\epsilon,\qquad \sigma_{+i}\epsilon=0.
\end{align}
In addition, if we assume the BPS solutions are static, and choose a gauge so that $A_\mu=(0, A_i)$, then the supersymmetry transformations of the fermionic fields in \eqref{delta-4} become
\begin{align}
&\delta_\epsilon\chi_a= i\sigma^1\bar\epsilon (D_1 - iD_2)\varphi_a -im_{ab} \epsilon\bar\varphi_b,\nn\\
	&\delta_\epsilon\lambda=i\epsilon F_{12}+ig\Big(|\varphi_1|^2-|\varphi_2|^2-\frac{\xi}g\Big)\epsilon.
\end{align}
The BPS equations, which are obtained by setting these variations to zero, are given by
\begin{align}
(D_1 - iD_2)\varphi_a=0,\qquad m_{ab}\varphi_b=0,\qquad B+g\Big(|\varphi_1|^2-|\varphi_2|^2-\frac{\xi}g\Big)=0. 
\end{align}
However, there exists no non-trivial configuration $(\varphi_a \neq 0)$ that simultaneously satisfies all the BPS equations derived above. 
This implies that the massive Abelian Higgs model does not admit any  BPS configuration.
Consequently, the only possible way to obtain non-constant solutions 
to the vacuum equations is to solve directly the second-order 
equations of motion given in \eqref{EOMm}. 
In general, these coupled nonlinear differential equations 
are analytically intractable. 
Therefore, constructing spatially dependent configurations 
that minimize the energy in the massive model remains a highly 
nontrivial problem.
In the next subsection, we instead turn to the massless Abelian Higgs model 
and investigate the existence of possible BPS configurations.

\subsection{Massless Abelian Higgs model}
The massless 4-dimensional $ {\cal N}=1$ supersymmetric Abelian Higgs model can include any number of chiral multiplets with different $U(1)$ charges. For definiteness, let us consider a model with two chiral multiplets with $U(1)$ charges $g_1$ and $g_2$. Then, the field content is, a massless vector multiplet $(A_\mu,\lambda)$ and two massless chiral multiplets $(\phi_a,\psi_a),~{\rm with}~a=1,2 $.  The Lagrangian of the  model is  \cite{Kim:2023abp}
\begin{align} \label{comfld2}
  {\cal L}& = -D_m\bar\phi_1D^m\phi_1-D_m\bar\phi_2D^m\phi_2-i\bar\psi_1\bar\sigma^m D_m\psi_1-i\bar\psi_2\bar\sigma^m D_m\psi_2-ig_1(\lambda\psi_1\bar\phi_1-\bar\lambda\bar\psi_1\phi_1)\nn\\
  &-ig_2(\lambda\psi_2\bar\phi_2-\bar\lambda\bar\psi_2\phi_2)-\frac1{4} F_{mn}F^{mn}-\frac i2\bar\lambda\bar\sigma^m \partial_m\lambda-\frac12\big(g_1|\phi_1|^2+g_2|\phi_2|^2-\xi\big)^2,
 \end{align}
 where the covariant derivatives of the matter fields are given by
\begin{align}\label{covder}
	D_m \begin{pmatrix}
		\phi_a \\ \psi_a \end{pmatrix} = \big[\partial_{m} + i g_a A_m \big] \begin{pmatrix} \phi_a \\ \psi_a \end{pmatrix}.
\end{align}
The Lagrangian is invariant under the ${\cal N}=1$ supersymmetry transformation:
\begin{align} \label{susyOS}
&\delta A_m=-\frac{i}{2}(\bar\epsilon\bar\sigma_m\lambda-\bar\lambda\bar\sigma_m\epsilon),\quad\delta\lambda=-\sigma^{mn}\epsilon F_{mn}+i\epsilon\big(g_1|\phi_1|^2+g_2|\phi_2|^2-\xi\big), 
\nn\\
&\delta\phi_a=\epsilon\psi_a, \quad \delta\psi_a =i\sigma^m\bar\epsilon D_m\phi_a. 
\end{align}

The null-reduction of Lagrangian in \eqref{comfld2}, with $\phi_a=e^{iMx^-}\varphi_a$ and $\psi_a=e^{iMx^-}\chi_a$ for both $a=1,2$, gives the non-relativistic Lagrangian, which is given by
\begin{align}\label{NRL}
	{\cal L}=&~~\frac12(\partial_t N)^2+\eta^{ij}E_i\partial_j N-\frac14 F_{ij}F^{ij}-\frac i2\bar\lambda\bar\sigma_- \partial_t\lambda-\frac i2\bar\lambda\bar\sigma^i \partial_i\lambda+i(M+g_aN)(\bar\varphi_a D_t\varphi_a-\bar D_t\bar\varphi_a \varphi_a)\nn\\
&-\bar D_i\bar\varphi_a D^i\varphi_a+(M+g_aN)\bar\chi_a\bar\sigma_+\chi_a-i\bar\chi_a\bar\sigma_- D_t\chi_a-i\bar\chi_a\bar\sigma^i D_i\chi_a\nn\\
&-ig_a\big(\lambda\chi_a\bar\varphi_a-\bar\lambda\bar\chi_a\varphi_a\big)-\frac{1}2 \big(g_1|\varphi_1|^2+g_2|\varphi_2|^2-\xi\big)^2.
\end{align}
Similarly, the null-reduction of the supersymmetry transformations of the fields gives
\begin{align}\label{delta-5}
	&\delta_\epsilon\varphi_a=\epsilon\chi_a, \qquad \delta_\epsilon\chi_a= -\big(M+g_aN\big)\sigma_+\bar\epsilon\varphi_a+i\sigma_-\bar\epsilon D_t\varphi_a+i\sigma^i\bar\epsilon D_i\varphi_a ,\nn\\
	&\delta_\epsilon\lambda=-2\sigma_{-+}\epsilon\partial_t N+2\sigma_{+i}\epsilon\partial_i N-2\sigma_{-i}\epsilon E_i-\sigma^{ij}\epsilon F_{ij}+i\big(g_1|\varphi_1|^2+g_2|\varphi_2|^2-\xi\big)\epsilon,\nn\\
	&\delta_\epsilon A_\mu=-\frac {i}2(\bar\epsilon\bar\sigma_\mu\lambda-\bar\lambda\bar\sigma_\mu\epsilon),\qquad \delta_\epsilon N=-\frac {i}2(\bar\epsilon\bar\sigma_-\lambda-\bar\lambda\bar\sigma_-\epsilon).
\end{align} 
In addition, the non-relativistic Lagrangian \eqref{NRL} is invariant under the bosonic transformations in \eqref{NRTrans} and \eqref{NRLrnz}, which are the space-time translations, the Galilean boost, space rotations, and a global $U(1)$ phase shift.    

The corresponding Hamiltonian can be obtained by calculating the conjugate momenta of the various fields from the Lagrangian in \eqref{NRL}. The result is
\begin{align}
	{\cal H}&=2g_a(M+g_aN)|\varphi_a|^2 A_t+\bar D_i\bar\varphi_a D^i\varphi_a-\sqrt2(M+g_aN)\bar\chi_{a\dot 2}\chi_{a2}+\sqrt2g_a\bar\chi_{a\dot 1}\chi_{a1}A_t\nn\\
	&-i\bar\chi_{a\dot 1}(D_1-iD_2)\chi_{a2}-i\bar\chi_{a\dot 2}(D_1+iD_2)\chi_{a1}-\frac i2\bar\lambda_{\dot 1}(\partial_1-i\partial_2)\lambda_2-\frac i2\bar\lambda_{\dot 2}(\partial_1+i\partial_2)\lambda_1+\frac12(\partial_t N)^2\nn\\
	&-\nabla^2 N A_t+\frac14 F_{ij}F^{ij}+ig_a\big(\lambda\chi_a\bar\varphi_a-\bar\lambda\bar\chi_a\varphi_a\big)+\frac{1}2 \big(g_1|\varphi_1|^2+g_2|\varphi_2|^2-\xi\big)^2.
\end{align}
To find the equilibrium states of this Hamiltonian, again we set the fermionic fields to zero and integrate out the Lagrange multiplier $A_t$, which imposes the Gauss-law constraint $\nabla^2N=2g_a(M+g_aN)|\varphi_a|^2$, then the Hamiltonian density reduces to
\begin{align}\label{energy-2}
	{\cal H}=|D_i\varphi_1|^2+|D_i\varphi_2|^2+\frac12(\partial_t N)^2+\frac12\big(g_1|\varphi_1|^2+g_2|\varphi_2|^2-\xi\big)^2+\frac12 B^2.
\end{align}
Varying the above  Hamiltonian density ${\cal H}$ with respect to all the fields, we obtain
\begin{align}\label{EOMml}
	&\delta_{\varphi_a}{\cal H}=\left[-\bar D_i^2+{g_a}\Big(g_1|\varphi_1|^2+g_2|\varphi_2|^2-\xi\Big)\right]\bar\varphi_a=0,\nn\\
	&\frac12\delta_{A_i}{\cal H}=g_a^2|\varphi_a|^2{\bf A}+g_a|\varphi_a|^2{\bf\nabla}\Omega_a+\frac12\nabla\times{\bf B}=0,
	\nn\\
	&\delta_{N}{\cal H}=-\partial_t^2N=0,
\end{align}
where again we have used $\varphi_a=|\varphi_a|e^{i\Omega_a}$.

\subsubsection{Constant extremum points}
In the case of the trivial solution $\varphi_1=\varphi_2=0$, the magnetic field should still satisfy $\nabla\times{\bf B}=0$,  which allows $\bf{B}$ to be represented as the gradient of a magnetic scalar potential. While a wide range of configurations can satisfy this depending on the boundary conditions, we consider the case where the field is a constant vector due to distant sources. 
Consequently, by assuming a uniform background magnetic field ${\bf B}={\bf H}$, the Hamiltonian density of the trivial solution is still
\begin{align}\label{symvac1}
	{\cal H}_{0}=\frac{\xi^2}{2}+\frac12H^2.
\end{align}

From the equations of motion in \eqref{EOMml}, one can construct constant, non-trivial solutions satisfying 
\(
|\varphi_a| \neq 0
\).
The existence of such solutions depends on the sign of the parameter~\(\xi\): 
one branch arises for \(\xi > 0\), while a distinct branch appears for \(\xi < 0\). 
For such constant configurations, the Ampère–Maxwell equation 
for the gauge field takes the form
\begin{align}
\frac{1}{2}\,\nabla \times \mathbf{B}
= - g_a^2 |\varphi_a|^2 \mathbf{A}
  - g_a |\varphi_a|^2 \nabla \Omega_a .
\end{align}
Since $|\varphi_a|$ is spatially constant, this equation reduces 
to the London equation. Consequently, in the bulk of the space, 
the magnetic field and the vector potential vanish, $\mathbf{B} = 0$ and $ \mathbf{A} = 0 $.
The constant non-trivial vacuum configuration is therefore 
characterized by
\begin{align}
\label{MLvac}
\frac{g_1 |\varphi_1|^2}{\xi}
+
\frac{g_2 |\varphi_2|^2}{\xi}
= 1 .
\end{align} 
For these non-trivial vacuum solutions, the Hamiltonian density becomes
\begin{align} {\cal H}_{vac}=0. \end{align} 
We note that, in the case of the massless model, since the energy of the non-trivial vacuum solutions is zero and that of the trivial vacuum is positive, the non-trivial vacuum solutions are always the favourable lowest energy stable vacuum states. We present the detailed vacuum moduli space in section \ref{sec4}.

\subsection{BPS equations}\label{BPSMSL}
The BPS equations of the massless supersymmetric Abelian Higgs model
can be derived from the Killing spinor equations,
following the same procedure outlined in subsection~\ref{BPSEQ}.
The resulting first-order differential equations and constraint are given by 
\begin{align}\label{BPSML}
(D_1 - i D_2)\varphi_1 = 0,
\qquad
(D_1 - i D_2)\varphi_2 = 0,
\qquad
B + \big(g_1 |\varphi_1|^2 + g_2 |\varphi_2|^2 - \xi \big) = 0 ,
\end{align}
where $B=F_{12}$. For later convenience, we note that under the reflection $x^2\to-x^2$ of the spatial coordinates, which flips $D_2\to-D_2$ and $F_{12}\to-F_{12}$, these equations take the holomorphic form
\begin{align}\label{BPSMLh}
(D_1 + i D_2)\varphi_1 = 0,
\qquad
(D_1 + i D_2)\varphi_2 = 0,
\qquad
B + \big(g_1 |\varphi_1|^2 + g_2 |\varphi_2|^2 - \xi \big) = 0 ,
\end{align}
where the magnetic field is now written as $B=\varepsilon^{ij}\partial_iA_j$ with $\varepsilon^{12}=-1$. In section~\ref{sec4}, we work with \eqref{BPSMLh} in the reflected coordinates.

It is worth emphasizing that, in a purely bosonic Abelian Higgs model,
where supersymmetry is absent, the BPS equations can only be obtained
through the Bogomol'nyi completion of the Hamiltonian.
In that case, the Hamiltonian can be rewritten as a sum of complete squares
plus a boundary term only if the scalar self-interaction potential
is tuned to the critical (self-dual) coupling,
which is fixed in terms of the gauge coupling constant.
By contrast, in supersymmetric theories,
the relation between the scalar self-coupling and the gauge coupling
is automatically dictated by supersymmetry.
Consequently, the BPS equations follow directly from the Killing spinor
conditions, obtained by setting the fermionic supersymmetry variations to zero.

In the absence of the mass term, the BPS equations in \eqref{BPSML}
admit non-trivial solutions for appropriate choices of the model parameters.
In the next section, we determine the necessary conditions
for the existence of such configurations
and analyse them using both analytic
and numerical methods.

\section{BPS Equations and Vortex Solutions}\label{sec4}
\subsection{Classification of vortex solutions}
As we have argued in subsection \ref{BPSEQ}, the massive Abelian Higgs model does not admit any non-trivial solutions to the vortex-type BPS equations.  We therefore restrict our attention to the massless theory, whose BPS equations are given in~\eqref{BPSMLh}. We note that, the asymptotic condition \eqref{MLvac} for vortex configurations admits a useful
topological interpretation in terms of the fractional vortices discussed in
\cite{Eto:2009bz}.  In this subsection, we adapt the argument of
\cite{Eto:2009bz} to the present setup, and classify the vortex solutions in a general setting. In the subsequent two subsections, we impose radial symmetry, and then explicitly find the vortex solutions.

We first parametrize the gauge-coupling parameters $g_a$ $(a=1,2)$ appearing in
the covariant derivative \eqref{covder} in terms of integer $U(1)$ charges
$q_a$ as
\begin{align}
    g_a = g_0 q_a,
    \qquad
    q_a \in \mathbb{Z}_{>0},
    \qquad
    \gcd(q_1,q_2)=1 .
    \label{norch}
\end{align}
Here $g_0$ denotes the minimal unit of charge, while $\gcd(q_1,q_2)$ is the
greatest common divisor of $q_1$ and $q_2$.  
The coprimality condition $\gcd(q_1,q_2)=1$ entails no loss of generality:
any common divisor of the two charges can be absorbed into the definition of
the unit charge $g_0$.

To classify vortex solutions, we first determine the vacuum moduli space in the
asymptotic region.  For the two charged scalar fields
$(\varphi_1,\varphi_2)$, the vacuum condition \eqref{MLvac} becomes
\begin{equation}
    q_1 |\varphi_1|^2 + q_2 |\varphi_2|^2
    =
    \frac{\xi}{g_0},
    \label{wvac}
\end{equation}
where we have used the parametrization \eqref{norch}.
The corresponding $U(1)$ gauge transformation in our setup acts as
\begin{align}\label{U(1)}
    \varphi_a
    &\longrightarrow
    e^{i q_a \alpha(x)} \varphi_a ,
    \nonumber \\
    A_\mu
    &\longrightarrow
    A_\mu - \frac{1}{g_0}\partial_\mu \alpha(x) ,
\end{align}
where $\alpha(x)$ is the gauge function.  Since field configurations related
by this gauge transformation are physically equivalent, the vacuum moduli space
is obtained by quotienting the vacuum manifold \eqref{wvac} by the
$U(1)$ action \eqref{U(1)}.  Thus one finds that the  vacuum moduli space is the weighted projective
space, 
\begin{equation}
    {\cal M}_{\rm vac}
    =
    \frac{
    \left\{
    (\varphi_1,\varphi_2)\in \mathbb{C}^2
    \,\big|\,
    q_1|\varphi_1|^2+q_2|\varphi_2|^2=\xi/g_0
    \right\}
    }{U(1)}
    \simeq
    W\mathbb{CP}^{1}_{(q_1,q_2)} .
\end{equation}
Equivalently, this space can be represented as a quotient by the complexified gauge group,
\begin{equation}\label{Cstar}
    (\varphi_1,\varphi_2)
    \sim
    \left(
    \lambda^{q_1}\varphi_1,
    \lambda^{q_2}\varphi_2
    \right),
    \qquad
    \lambda \in \mathbb{C}^{*},
\end{equation}
where $\mathbb{C}^{*} \equiv \mathbb{C}\setminus\{0\}$ denotes the set of nonzero complex numbers.

In the special case $q_1=q_2=1$, the weighted projective space reduces to the
ordinary complex projective space,
\begin{equation}
    W\mathbb{CP}^{1}_{(1,1)}
    \simeq
    \mathbb{CP}^{1}.
\end{equation}
This is the familiar complex projective line.  Just as $\mathbb{CP}^1$ can be
constructed by gluing the two coordinate patches $\varphi_1\neq 0$ and
$\varphi_2\neq 0$, the weighted projective space
$W\mathbb{CP}^{1}_{(q_1,q_2)}$ is also covered by two corresponding patches.
As a topological space, $W\mathbb{CP}^{1}_{(q_1,q_2)}$ is homeomorphic to
$\mathbb{CP}^1$, and hence to the two-sphere $S^2$.  However, its local quotient
structure is in general different from that of the ordinary projective space.
In particular, for nontrivial weights $(q_1,q_2)$, the space carries orbifold
singularities at the points where one of the complex fields vanishes and the other field has charge $q_a>1$.

Let us make this statement more explicit. For a generic point with $\varphi_1\neq 0$ and $\varphi_2\neq 0$, invariance under the 
$\mathbb{C}^*$-action \eqref{Cstar} requires
\begin{equation}
    \lambda^{q_1}=1,
    \qquad
    \lambda^{q_2}=1 .
\end{equation}
Since we have assumed $\gcd(q_1,q_2)=1$, the only common solution is $\lambda=1$. 
Thus the stabilizer is trivial at a generic point, and the quotient is locally smooth there.
On the other hand, consider the point with $\varphi_1=0$ and $\varphi_2\neq 0$. 
Under the $\mathbb{C}^*$-action \eqref{Cstar}, this point transforms as
\begin{equation}
    (0,\varphi_2)
    \sim
    (0,\lambda^{q_2}\varphi_2).
\end{equation}
It remains invariant if and only if
\begin{equation}
    \lambda^{q_2}=1 .
\end{equation}
The solutions are
\begin{equation}
    \lambda
    =
    e^{2\pi i k/q_2},
    \qquad
    k=0,1,\ldots,q_2-1 ,
\end{equation}
which form the discrete group $\mathbb{Z}_{q_2}$. 
Therefore, a residual discrete symmetry $\mathbb{Z}_{q_2}$ remains at $\varphi_1=0$, and this point is a 
$\mathbb{Z}_{q_2}$ fixed point.
Similarly, at the point with $\varphi_1\neq 0$ and $\varphi_2=0$, the residual discrete symmetry is 
$\mathbb{Z}_{q_1}$. 
Hence this point is a $\mathbb{Z}_{q_1}$ fixed point. 
Consequently, the special points $(0,\varphi_2)$ and $(\varphi_1,0)$ are locally modeled on the orbifold quotients
\begin{equation}
    \mathbb{C}/\mathbb{Z}_{q_2}
    \qquad \text{and} \qquad
    \mathbb{C}/\mathbb{Z}_{q_1},
\end{equation}
respectively.
Hence the fixed point $(0,\varphi_2)$ (or $(\varphi_1,0)$) is an orbifold singularity of
$W\mathbb{CP}^1_{(q_1,q_2)}$ for $q_2>1$ (or $q_1>1$), whereas for $q_a=1$ the corresponding point is an ordinary smooth point.


The orbifold singularities of the vacuum moduli space
$W\mathbb{CP}^{1}_{(q_1,q_2)}$ modify the usual quantization condition for
vortex winding numbers.  In particular, when the asymptotic value of the scalar
fields lies at an orbifold fixed point, the corresponding vortex can carry a
fractional winding number.  To make this point explicit, we use the
moduli-matrix formalism, in which the origin of the fractional quantization is
particularly transparent.  Several numerical BPS vortex solutions with
fractional winding numbers will be presented in subsection~\ref{nsol}.

The gauged Cauchy--Riemann equations in \eqref{BPSMLh} can be solved
by introducing a complex function $s=s(z,\bar z)$.
In complex coordinates\footnote{
Here we use complex coordinates $z \equiv x+iy$, $\bar z \equiv x-iy$, and define 
\begin{align}
A_z \equiv \frac{1}{2}\left(A_1-iA_2\right),
    \qquad
    A_{\bar z} \equiv \frac{1}{2}\left(A_1+iA_2\right).
\end{align}}
one may write the gauge field in terms of a complex function $s = s(z, \bar z)$ as
\begin{align}
A_{\bar{z}} (z, \bar z) =
    -\,\frac{i}{g_0}\,\partial_{\bar{z}} \log s(z, \bar z).
\end{align}
Then, the gauged Cauchy--Riemann equation is reduced to
\begin{equation}
    \partial_{\bar{z}}\!\left(s^{q_a}\varphi_a\right)=0 .
\end{equation}
Therefore, we obtain 
\begin{equation}
    \varphi_a (z, \bar z)
    =
    s^{-q_a}\varphi_{a,0}(z),
    \qquad
    \partial_{\bar z}\varphi_{a,0}(z)=0 ,
\end{equation}
where $\varphi_{a,0}(z)$'s are called the holomorphic moduli matrix functions.

For a ${\rm U}(1)$ winding number $\nu$, the complex function $s(z, \bar z)$ in the asymptotic limit $|z|\to \infty$ behaves as
\begin{align}
s(z, \bar{z}) \sim z^\nu.
\end{align}
Then, the holomorphic moduli matrix functions $\varphi_{a,0}(z)$ in the asymptotic limit behave as 
\begin{align}\label{homo}
\varphi_{a,0} (z) \sim z^{q_a \nu}. 
\end{align}

At a generic point of the vacuum moduli space, where both asymptotic
vacuum expectation values are nonvanishing, the holomorphy condition
in \eqref{homo} requires
\begin{equation}\label{windnum1}
    q_1 \nu \in \mathbb{Z}_{\geq 0},
    \qquad
    q_2 \nu \in \mathbb{Z}_{\geq 0}.
\end{equation}
Since \(q_1\) and \(q_2\) are relatively prime, the minimal generic
solutions are given by
\begin{equation}
    \nu = k,
    \qquad
    k \in \mathbb{Z}_{>0}.
\end{equation}
It then follows from the asymptotic behavior of the two complex scalar
fields in \eqref{homo} that their phases wind as
\begin{equation}
    n_1 = q_1 k,
    \qquad
    n_2 = q_2 k .
\end{equation}
Using the relation \(g_a = g_0 q_a\), we immediately obtain
\begin{equation}
    \frac{n_1}{g_1}
    =
    \frac{q_1 k}{g_0 q_1}
    =
    \frac{q_2 k}{g_0 q_2}
    =
    \frac{n_2}{g_2},
\end{equation}
which is precisely the condition \eqref{ProtCond}. Hence,
\eqref{ProtCond} is the combined finite-energy and holomorphy condition
requiring the two scalar fields to wind along the same broken
\(U(1)\) gauge orbit at spatial infinity.

On the other hand, at an orbifold fixed point, for instance when
\(\varphi_1 = 0\) and \(\varphi_2 \neq 0\) in the asymptotic limit,
only the condition
\begin{equation}
    q_2 \nu \in \mathbb{Z}_{\geq 0}
\end{equation}
is required by \eqref{homo} and \eqref{windnum1}. Therefore, if the
asymptotic vacuum of the vortex solution lies at this orbifold fixed
point, the minimal winding number is $\nu = 1/q_2$. 
Similarly, a vortex solution whose asymptotic vacuum satisfies
\(\varphi_1 \neq 0\) and \(\varphi_2 = 0\) has the minimal winding number $\nu = 1/q_1$. 
Consequently, at special orbifold points of
\(W\mathbb{CP}^{1}_{(q_1,q_2)}\), one of the two scalar vacuum expectation
values vanishes, and the allowed \(U(1)\) winding can become fractional.
More generally, the allowed fractional windings are of the form
\begin{equation}
    \nu = \frac{k}{q_1}
    \qquad \text{or} \qquad
    \nu = \frac{k}{q_2},
    \qquad
    k \in \mathbb{Z}_{>0},
\end{equation}
depending on which orbifold fixed point is approached at spatial
infinity. These configurations are the fractional vortices discussed in
\cite{Eto:2009bz}.

 \subsection{Radially symmetric solutions}

 To determine the necessary conditions for the existence of nontrivial vortex-type solutions in which both complex scalar fields $\varphi_a$ $(a=1,2)$ are nonvanishing, we assume radial symmetry. 
 In polar coordinates $(r,\theta)$, the scalar and gauge fields are parametrized as
 \begin{align}\label{radans}
 	\varphi_a &= |\varphi_a|(r)e^{i n_a \theta}, \nonumber \\
 	A^i &= - \varepsilon^{ij}x_j\frac{A(r)}{r^2}
 	= \left(\frac{\sin\theta}{r}A(r), - \frac{\cos\theta}{r}A(r)\right).
 \end{align}
 To obtain vortex solutions of the BPS equations in \eqref{BPSMLh}, the winding numbers $n_a$ are required to be non-negative integers.
 Substituting the ansatz in \eqref{radans} into the BPS equations in \eqref{BPSMLh}, we obtain
 \begin{align}\label{BPSrad}
 	&\cos\theta 
 	\Big(
 	\frac{d}{dr}
 	+\frac{g_aA-n_a}{r}
 	\Big)
 	|\varphi_a|
 	+i\sin\theta
 	\Big(
 	\frac{d}{dr}
 	+\frac{g_aA-n_a}{r}
 	\Big)
 	|\varphi_a|
 	=0,
 	\nn\\
 	&\frac{1}{r}\frac{dA}{dr}
 	+\left(
 	g_1|\varphi_1|^2
 	+g_2|\varphi_2|^2
 	-\xi
 	\right)=0.
 \end{align}
 Requiring both the real and imaginary parts of the first equation in \eqref{BPSrad} to vanish independently leads to
 \begin{align}\label{A(r)}
 	\frac{d|\varphi_1|}{dr}
 	=
 	\frac{n_1 - g_1A}{r}|\varphi_1|,
 	\qquad
 	\frac{d|\varphi_2|}{dr}
 	=
 	\frac{n_2 - g_2A}{r}|\varphi_2|.
 \end{align}
 
 For finite-energy solutions approaching nonzero constant vacuum values satisfying \eqref{MLvac} in the asymptotic region, the gauge field must satisfy
 \begin{align}\label{AsyA}
 	A(r\to\infty)
 	=
 	\frac{n_1}{g_1}
 	=
 	\frac{n_2}{g_2}.
 \end{align}
 Therefore, nontrivial vortex solutions with both scalar fields nonvanishing exist only when the $U(1)$ charges satisfy
 \begin{align}\label{ProtCond}
 	\frac{n_1}{g_1}
 	=
 	\frac{n_2}{g_2},
 	\qquad
 	n_1,\,n_2\in \mathbb{Z}_{>0}.
 \end{align}
 This condition implies that $g_1$ and $g_2$ cannot have opposite signs. 
 The same argument can be straightforwardly generalized to models with more than two complex scalar fields.

 Assuming that the above necessary condition is satisfied, next we present the detailed procedure of finding the radially symmetric solutions to our BPS equations. Inserting $\varphi_a=|\varphi_a|e^{i\Omega_a}$, both the real and the imaginary parts of the first two equations in \eqref{BPSMLh}, gives us
 \begin{align}
 \partial_i\ln|\varphi_a| + \varepsilon^{ij}\big(g_{a} A_j+\partial_j\Omega_a\big)=0.
 \end{align}
 Applying one additional partial derivative $\partial_i$ throughout this equation, we obtain
 \begin{align}\label{DBPS}
 \nabla^2\ln|\varphi_a|+g_aB-2\pi\sum_{m=1}^{n_a}\delta^{(2)}({\bf x}-{\bf x}_m)=0,
 \end{align}
 where we have replaced 
 \begin{align}
 B=\varepsilon^{ij}\partial_i A_j,\qquad \varepsilon^{ij}\partial_i \partial_j\Omega_a=\varepsilon^{ij}\partial_i \partial_j(n_a\theta)=-2\pi\sum_{m=1}^{n_a}\delta^{(2)}(\bf{x}-\bf{x}_m). 
 \end{align}
 Eliminating $B$ from \eqref{DBPS} by using the third equation in \eqref{BPSMLh}, we can write 
 \begin{align}\label{DBPSb}
 \nabla^2\ln|\varphi_a|^2-4\pi \sum_{m=1}^{n_a}\delta^{(2)}({\bf x}-{\bf x}_m)=2g_a\big(g_1|\varphi_1|^2+g_2|\varphi_2|^2-\xi\big),
 \end{align}
 For ${\bf x}\ne{\bf x}_m$ the delta function terms vanish, and imposing the rotational symmetry on both $\varphi_1$ and $\varphi_2$, we obtain 
  \begin{align}\label{vortex1}
  &\left(\frac{d^2}{dr^2}+\frac1r\frac d{dr}\right)\ln|\varphi_1|=g_1\left(g_1|\varphi_1|^2+g_2|\varphi_2|^2-\xi\right),\nn\\
  &\left(\frac {d^2}{dr^2}+\frac1r\frac d{dr}\right)\ln|\varphi_2|=g_2\left(g_1|\varphi_1|^2+g_2|\varphi_2|^2-\xi\right).
  \end{align}
  
  Now, we consider the asymptotic solutions.
  First, we consider the case in which both asymptotic values are nonzero.
  Let us make a solution ansatz $|\varphi_a|(r)=|\varphi_a^0|-C_af_a(r)$, where $|\varphi_a^0|$ are the constant vacuum solutions obtained in the previous section, whereas  the additional terms $C_af_a(r)\ll |\varphi_a^0|$ are real valued and they define small deviations from those constant vacuum solutions. Inserting this ansatz in \eqref{vortex1} and keeping terms only up to linear in ${\cal O}\left(\frac{C_af_a}{|\varphi_a^0|}\right)$, we obtain
  \begin{align}\label{vortex2}
  &\left(\frac{d^2}{dr^2}+\frac1r\frac d{dr}\right)\begin{pmatrix}
	f_1\\
	f_2
\end{pmatrix}=L\begin{pmatrix}
	f_1\\
	f_2
\end{pmatrix}
  \end{align}
  where we have used the constant vacuum equation $\big(g_1|\varphi_1^0|^2+g_2|\varphi_2^0|^2-\xi =0\big)$, and also introduced the matrix
  \begin{align}
 L=2\begin{pmatrix}
	g_1^2|\varphi_1^0|^{2} & g_1g_2\frac{C_2}{C_1}|\varphi_1^0| |\varphi_2^0|\\
	g_1g_2\frac{C_1}{C_2} |\varphi_{1}^{0}| |\varphi_2^0| &g_2^2|\varphi_2^0|^{2} 
\end{pmatrix}.
  \end{align}
  The eigenvalues of this matrix are $\big(0,\ell^2\big)$, where $\ell^2=2g_1^2|\varphi_1^0|^{2} + 2g_2^2|\varphi_2^0|^{2}$.
  Then the diagonalized equations obtained from \eqref{vortex2} are given by
  \begin{align}\label{vortex3}
 \frac{d^2\tilde f_1}{dr^2}+\frac1r\frac {d\tilde f_1}{dr}=0,\qquad \frac{d^2\tilde f_2}{dr^2}+\frac1r\frac {d\tilde f_2}{dr}=\ell^2\tilde f_2.
  \end{align} 
  The general solution to the first equation in \eqref{vortex3} is $\tilde f_1=a_1\ln r+a_2$. If $a_1\ne0$, this solution is asymptotically divergent, hence, we have to choose $a_1=0$. Furthermore, $a_2\to 0$, in order to agree with the asymptotic solution $|\varphi_a^0|$. Therefore, the acceptable solution is $\tilde f_1=0$.   The solution to the second equation in \eqref{vortex3} is given in terms of the modified Bessel function as $\tilde f_2(r)=K_0( \ell r)$. Then, using  the matrix that diagonalizes $L$, we obtain the original functions, which are given by
  \begin{align}
  f_1(r)=-\sin\beta~K_0(\ell r), \qquad f_2(r)=\cos\beta~K_0(\ell r), 
  \end{align}
  where $\beta$ is defined by $\tan \beta = - \frac{C_{2}}{C_{1}} \frac{g_{1} |\varphi_{1}^{0}|}{g_{2} |\varphi_{2}^{0}|}$. 
  
  Now we can use the asymptotic solutions 
  \begin{align}
  |\varphi_1|=|\varphi_1^0|+C_1\sin\beta~K_0(\ell r), \qquad |\varphi_2|=|\varphi_2^0|-C_2\cos\beta~K_0(\ell r)
  \end{align}
  in \eqref{A(r)}, to find the asymptotic solution for $A(r)$, which is given by
  \begin{align}\label{A(r)sol}
  A(r)=\frac{n_1}{g_1}+C_1\frac{\sin\beta}{g_1|\varphi_1^0|}(\ell r)K_1(\ell r)=\frac{n_2}{g_2}-C_2\frac{\cos\beta}{g_2|\varphi_2^0|}(\ell r)K_1(\ell r),
  \end{align}
  where we have used $\frac{dK_0(x)}{dx}=-K_{1} (x)$. Using \eqref{AsyA} and the definition of $\beta$, one can verify that the two solutions in \eqref{A(r)sol} are equivalent. The corresponding magnetic field is given by
  \begin{align}
  B=\frac1r\frac{dA}{dr}=-C_1\frac{\sin\beta}{g_1|\varphi_1^0|}\ell^2K_0(\ell r)=C_2\frac{\cos\beta}{g_2|\varphi_2^0|}\ell^2K_0(\ell r),
\end{align}   
  where we have used $\frac{dK_1(x)}{dx}=- K_0(x)-\frac1x K_1(x)$. Finally, using these asymptotic solutions in \eqref{energy-2}, we can write the asymptotic form of the Hamiltonian density as
  \begin{align}
  {\cal H}=C_1^2\frac{\sin^2\beta}{g_1^2 |\varphi_1^0|^2}\ell^4 \big[K_0^2(\ell r)+K_1^2(\ell r)\big]=C_2^2\frac{\cos^2\beta}{g_2^2 |\varphi_2^0|^2}\ell^4 \big[K_0^2(\ell r)+K_1^2(\ell r)\big]. 
  \end{align}
  
  Next, we consider the asymptotic behaviour of the solutions in which one of the scalar fields vanishes asymptotically.
  In the above discussion, we have considered the topological vortex solutions, that is we considered the case of non-zero asymptotic values for both complex scalars. A natural question is whether there exist other rotationally symmetric solutions, such as nontopological vortices or Q-balls, which are also nontopological solutions. Since the nontopological solutions vanish asymptotically, we can consider two possible behaviors as $r \to \infty$:
  \begin{align}
  	&|\varphi_{1}| = 0 = |\varphi_{2}|, \nonumber\\
  	&|\varphi_{1}| = 0 \neq |\varphi_{2}| ~ (\text{or } |\varphi_{1}| \neq 0 = |\varphi_{2}|).
  \end{align}
  However, the first case, which corresponds to a two Q-balls profile, $(|\varphi_{1}| = 0 = |\varphi_{2}|)$, does not satisfy the boundary condition \eqref{MLvac}.
  Therefore, the only possibility for a finite energy configuration is the second case, $|\varphi_{1}| = 0 \neq |\varphi_{2}|$ (or $|\varphi_{1}| \neq 0 = |\varphi_{2}|)$.

  Following the previous subsection, we can consider the second case, $|\varphi_{1}| = 0 \neq |\varphi_{2}|$ (or $|\varphi_{1}| \neq 0 = |\varphi_{2}|)$ with finite energy, whose winding is fractional when the surviving field carries a charge $q_a>1$.
  In this case, the asymptotic solution shows a power-law behavior, unlike the above topological vortices,
  \begin{align}
  	|\varphi_{1}| =&\, \sqrt{\frac{\xi}{g_{1}}} \Big( 1 - \frac{g_{2} C_{2}^{2}}{2 \xi} r^{-\epsilon} \Big) ,\nonumber\\
  	|\varphi_{2}| =&\, C_{2} r^{-\epsilon/2},
  \end{align}
  where $\epsilon > 0$ is determined below.

  As we did in the case of topological vortices, we can verify the boundary condition in \eqref{AsyA}. For this case,
  \begin{align}
  	A(r\to\infty) = \frac{n_{1}}{g_{1}} = \frac{n_{2} + \frac{\epsilon}{2}}{g_{2}}
  	.\label{ProtCond2}
  \end{align}
  This determines an analytic expression of $\epsilon$, which is
  \begin{align}
  	\epsilon = 2\Big( n_{1} \frac{g_{2}}{g_{1}} - n_{2} \Big)
  	.\label{epsilon}
  \end{align}
 
  \subsection{Numerical solutions} \label{nsol}

  The above asymptotic solutions to the BPS equation tell us the behaviour  of the solutions to the equations of motion in asymptotic regions. In order to get the full picture of the solutions, we need to use numerical methods to find the complete solutions. To that end, let us consider the system of coupled BPS equations for the scalar field $|\varphi_a|(r)$ and the gauge field $A(r)$, which can be read from \eqref{BPSrad} and they are given by   
\begin{align}
	\frac{d |\varphi_{a}|}{dr} =&\, \frac{n_{a} - g_{a} A}{r} |\varphi_a| ,\label{shw1}\\
	\frac{1}{r} \frac{dA}{dr} =&\, -\big( g_{1} |\varphi_{1}|^{2} + g_{2} |\varphi_{2}|^{2} - \xi \big) \label{shw2}.
\end{align}
By imposing the radial symmetry, these BPS equations were reduced to ordinary differential equations for superimposed multi-vortices. 
For non-singular BPS solutions, the boundary conditions at the origin are
\begin{align}
	|\varphi_{a}| (0) = 0, \qquad A(0) = 0,
\end{align}
and the finite energy requirement ensures the asymptotic boundary conditions should be
\begin{align}
	g_{1} |\varphi_{1}|^{2} (r \to\infty) + g_{2} |\varphi_{2}|^{2} (r \to\infty) = \xi , \qquad A(r \to\infty) = \frac{n_{1}}{g_{1}} = \frac{n_{2}}{g_{2}} .\label{shw3}
\end{align}
Under these boundary conditions, the magnetic flux is proportional to the vortex number,
\begin{align}
	\Phi_{B} = \int d^{2} x \, B = 2\pi \int dr \, r \frac{1}{r} \frac{dA}{dr} = 2\pi \big[ A(\infty) - A(0) \big] = \frac{2\pi n_{1}}{g_{1}} = \frac{2\pi n_{2}}{g_{2}}
	.
\end{align}
The energy for BPS solution is saturated to the Bogomolny bound as
\begin{align}
	E &\,= \int d^{2} x \, \bigg\{ |(D_{1} + i D_{2}) \varphi_{1}|^{2} + |(D_{1} + i D_{2}) \varphi_{2}|^{2} +\frac{1}{2} \Big[ B +  \big( g_{1} |\varphi_{1}|^{2} + g_{2} |\varphi_{2}|^{2} - \xi \big) \Big]^{2} \nonumber\\	&\,~~~~~~~~~~~~~~ +  \xi B \bigg\} \ge \xi \Phi_{B} = \Big(\frac{n_{1}}{g_1} + \frac{n_{2}}{g_2}\Big) \pi \xi.
\end{align}

\begin{figure}[H]
	\centering
	\subfigure[]{
		\includegraphics[height=0.25\textheight]{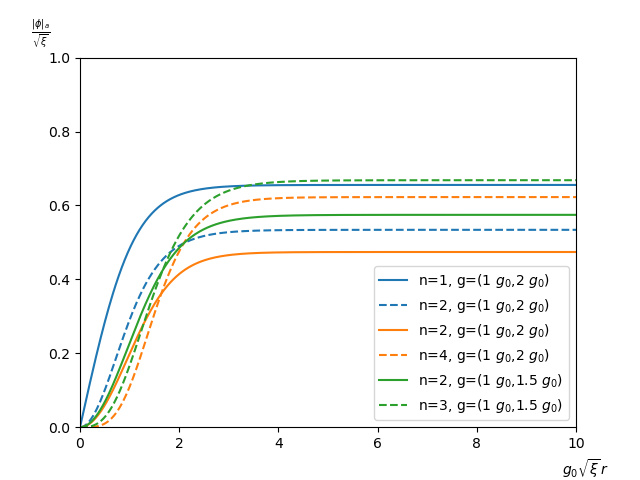}
	}
	\subfigure[]{
		\includegraphics[height=0.25\textheight]{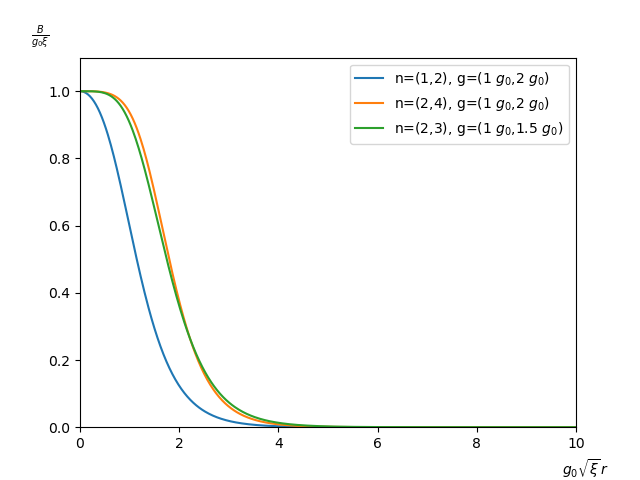}
	}
	\subfigure[]{
		\includegraphics[height=0.25\textheight]{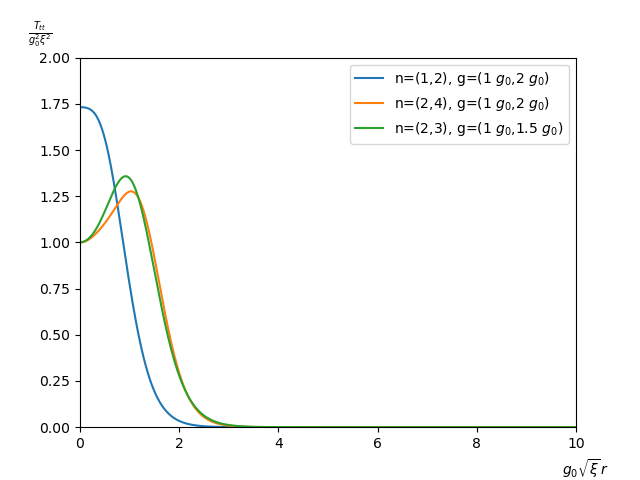}
	}
	\caption{Profiles of vortex solutions for different choices of gauge couplings and winding numbers. (a) Solution profiles as functions of $g_{0} \sqrt{\xi}\, r$, where the solid and dashed curves represent the $n_1$, $n_2$ vortex solutions, respectively. All profiles connect regularly at the origin and the asymptotic vacuum. (b) Magnetic field profiles, which are localized near the vortex core and decay exponentially at large distances. (c) Energy density distributions, which are concentrated around the vortex centers. The three representative cases shown are $(g_1,g_2,n_1,n_2) = (g_0, 2 g_0, 1,2)$, $(g_0, 2 g_0, 2,4)$, $(g_0, 1.5 g_0, 2,3)$.}
	\label{fig:1}
\end{figure}

By numerically solving the coupled BPS equations \eqref{shw1} and \eqref{shw2}, we obtain coupled vortex solution profiles for the scalar fields, magnetic fields, and energy densities. We consider several cases by varying the gauge coupling and the vortex number $(g_{1},g_{2}, n_{1}, n_{2}) = (g_{0}, 2 g_{0}, 1, 2)$, $(g_{0}, 2 g_{0}, 2, 4), (g_{0}, 1.5 g_{0}, 2, 3)$, where $g_{0}$ denotes a unit reference gauge coupling to fix the overall scale and those conditions satisfy the topological protection condition \eqref{ProtCond}. In the numerical computations, we set $g_0=\xi=1$. The plots of these numerical solutions are as shown in Figure \ref{fig:1}.

The scalar fields vanish at the origin and monotonically approach the vacuum values at large $r$, satisfying \eqref{shw3}. As the vortex number increases, the vortex size becomes larger, as shown in Figure \ref{fig:1}-(a).
The magnetic fields are localized near the zeros of vortex and decay exponentially at large distances, which is consistent with the London equations discussed in section 3. Increasing the vortex number enlarges the magnetic core, as shown in Figure \ref{fig:1}-(b).
The energy density profiles are centered at the zeros of vortex and exponentially decay to zero at large distance, as shown in Figure \ref{fig:1}-(c).
In all cases, the solutions are smooth, satisfy the boundary conditions \eqref{shw3}, and the total energy agrees with the BPS value $E =  \xi \Phi_{B} = \Big(\frac{n_{1}}{g_1} + \frac{n_{2}}{g_2}\Big) \pi \xi$.

Next we consider the nontopological solution cases. For non-singular BPS solutions, the boundary conditions at the origin are
\begin{align}
	|\varphi_{a}| (0) = |\varphi_{a}^{00}| \delta_{n_{a} 0}, \qquad A(0) = 0 ,
\end{align}
where $|\varphi_{a}^{00}|$ is nonzero value for nontopological profile.

For convenience, we assume that $|\varphi_{2}|$ goes to zero at the asymptote.
The finite energy requirement ensures the asymptotic boundary conditions should be
\begin{align}
	g_{1} |\varphi_{1}|^{2} (r \to\infty) = \xi , \qquad A(r\to\infty) = \frac{n_{1}}{g_{1}} = \frac{n_{2} + \frac{\epsilon}{2}}{g_{2}} .\label{shw4}
\end{align}
Under these boundary conditions, the magnetic flux is proportional to the vortex number,
\begin{align}
	\Phi_{B} = \int d^{2} x \, B = 2\pi \int dr \, r \frac{1}{r} \frac{dA}{dr} = 2\pi \big[ A(\infty) - A(0) \big] = \frac{2\pi n_{1}}{g_{1}} = \frac{2\pi (n_{2} + \frac{\epsilon}{2})}{g_{2}}
	.
\end{align}
The energy for BPS solution is saturated to the Bogomolny bound as
\begin{align}
	E &\,= \int d^{2} x \, \bigg\{ |(D_{1} + i D_{2}) \varphi_{1}|^{2} + |(D_{1} + i D_{2}) \varphi_{2}|^{2} +\frac{1}{2} \Big[ B +  \big( g_{1} |\varphi_{1}|^{2} + g_{2} |\varphi_{2}|^{2} - \xi \big) \Big]^{2} \nonumber\\	&\,~~~~~~~~~~~~~~ +  \xi B \bigg\} \ge \xi \Phi_{B} = \Big(\frac{n_{1}}{g_1} + \frac{n_{2} + \frac{\epsilon}{2}}{g_2}\Big) \pi \xi.
\end{align}

\begin{figure}[H]
	\centering
	\subfigure[]{
		\includegraphics[height=0.25\textheight]{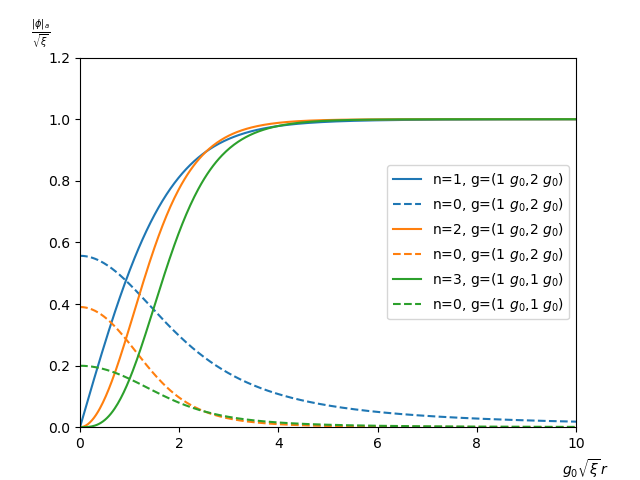}
	}
	\subfigure[]{
		\includegraphics[height=0.25\textheight]{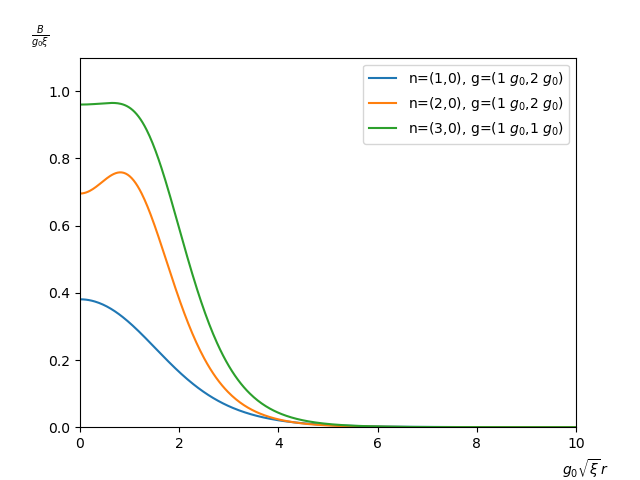}
	}
	\subfigure[]{
		\includegraphics[height=0.25\textheight]{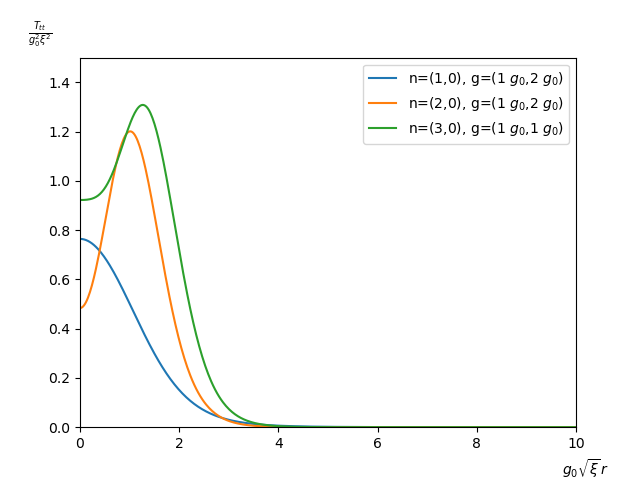}
	}
	\caption{Profiles of vortex solutions for different choices of gauge couplings and winding numbers. (a) Solution profiles as functions of $g_{0} \sqrt{\xi}\, r$, where the solid and dashed curves represent the $n_1$, $n_2$ solutions, respectively. All profiles connect regularly at the origin and the asymptotic vacuum. (b) Magnetic field profiles, which are localized near the vortex core and decay as power laws at large distance. (c) Energy density distributions, which are concentrated around the vortex centers. The three representative cases shown are $(g_1,g_2,n_1,n_2) = (g_0, 2 g_0, 1,0)$, $(g_0, 2 g_0, 2,0)$, $(g_0, g_0, 3,0)$. The numerical values of the exponents are $\epsilon = 4.00, 7.99, 5.99$, which are consistent with the condition \eqref{epsilon} up to numerical error.}
	\label{fig:2}
\end{figure}

By numerically solving the coupled BPS equations \eqref{shw1} and \eqref{shw2}, we obtain coupled vortex solution profiles for the scalar fields, magnetic fields, and energy densities. We consider several cases by varying the gauge coupling and the vortex number $(g_{1},g_{2}, n_{1}, n_{2}) = (g_{0}, 2 g_{0}, 1, 0)$, $(g_{0}, 2 g_{0}, 2, 0), (g_{0}, g_{0}, 3, 0)$, where $g_{0}$ denotes a unit reference gauge coupling to fix the overall scale and those conditions satisfy the topological protection condition \eqref{ProtCond2}. The plots of these numerical solutions are as shown in Figure \ref{fig:2}.
In all cases, the solutions are smooth, satisfy the boundary conditions \eqref{shw4}, and the total energy agrees with the BPS value $E =  \xi \Phi_{B} = \Big(\frac{n_{1}}{g_1} + \frac{n_{2} + \frac{\epsilon}{2}}{g_2}\Big) \pi \xi$.

\begin{figure}[H]
	\centering
	\subfigure[]{
		\includegraphics[height=0.25\textheight]{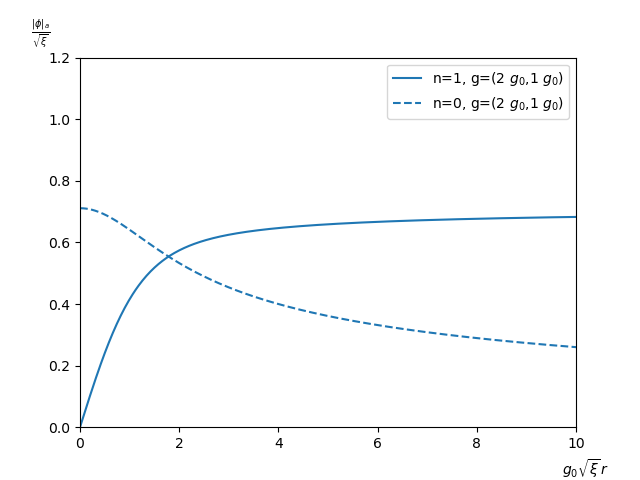}
	}
	\subfigure[]{
		\includegraphics[height=0.25\textheight]{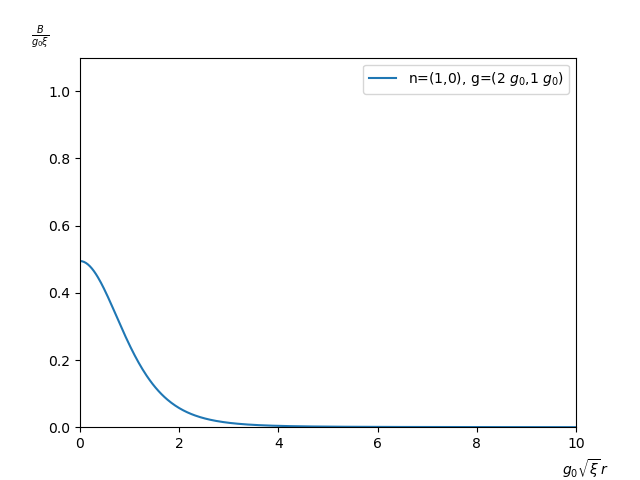}
	}
	\subfigure[]{
		\includegraphics[height=0.25\textheight]{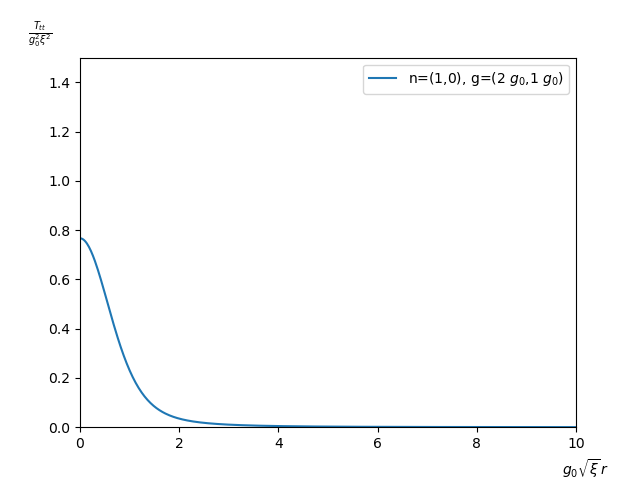}
	}
	\caption{Scalar profiles of the half-flux vortex solution with $(g_1,g_2,n_1,n_2)=(2g_0,g_0,1,0)$ and $C_2=0.5$, as functions of $g_{0} \sqrt{\xi}\, r$. The solid and dashed curves represent $\varphi_1$ and $\varphi_2$, respectively, and the dotted line indicates the asymptotic value $|\varphi_1|=\sqrt{\xi/2g_0}$. The field $\varphi_2$ decays as $r^{-1/2}$, and the solution carries the fractional flux $\Phi_B=\pi/g_0$.}
	\label{fig:3}
\end{figure}

For the three cases in Figure~\ref{fig:2}, the surviving field $\varphi_1$ carries the unit charge $q_1=1$, so that the asymptotic vacuum is an ordinary smooth point of the vacuum moduli space and the flux $\Phi_B=2\pi n_1/g_0$ is an integer multiple of $2\pi/g_0$, even though $\varphi_2$ decays algebraically. A fractional vortex requires the surviving field to carry a charge $q_a>1$. As the simplest example, we take $(g_1,g_2,n_1,n_2)=(2g_0,g_0,1,0)$, for which $|\varphi_1|\to\sqrt{\xi/2g_0}$ and $\varphi_2\to0$ asymptotically, i.e., the asymptotic vacuum is the $\mathbb{Z}_2$ orbifold point of $W\mathbb{CP}^1_{(2,1)}$. Then \eqref{ProtCond2} and \eqref{epsilon} give $\epsilon=1$, and
\begin{align}\label{halfflux}
	\nu=\frac{g_0\Phi_B}{2\pi}=\frac12,\qquad \Phi_B=\frac{\pi}{g_0},\qquad E=\xi\Phi_B=\frac{\pi\xi}{g_0},\qquad |\varphi_2|\simeq C_2\, r^{-1/2}.
\end{align}
The numerical solution with $C_2=0.5$ is shown in Figure~\ref{fig:3}. Because of the slow algebraic tail of $\varphi_2$, the flux and the energy evaluated within a finite radius $R$ approach the values in \eqref{halfflux} only as $1/r$.

\section{Conclusions}

In this work we studied the construction and soliton structure of non-relativistic supersymmetric gauge theories obtained through the method of null reduction. Starting from the relativistic ${\cal N}=1$ supersymmetric Abelian Higgs model in $(3+1)$ dimensions, we systematically applied null reduction to derive a non-relativistic theory in $(2+1)$ dimensions. This procedure preserves the essential symmetry structure of the relativistic parent theory, transforming the super-Poincar\'e algebra into the supersymmetric extension of the Bargmann algebra, which governs Galilean-invariant systems. As a result, the resulting theory maintains gauge invariance and supersymmetry while exhibiting the Galilean symmetry characteristic of non-relativistic systems.

Within this framework, we obtained the non-relativistic limits of massive as well as massless supersymmetric Abelian Higgs models and examined their vacuum structure and soliton solutions. In the massive model, the presence of the mass term together with the Fayet--Iliopoulos term leads to spontaneous supersymmetry breaking. By analyzing the Hamiltonian density and the corresponding extremum conditions, we determined the possible vacuum configurations and showed that the Killing spinor conditions do not allow non-trivial BPS configurations. Consequently, the massive theory does not admit vortex-type BPS solutions, and spatially dependent configurations would have to be obtained directly from the full second-order equations of motion, which are analytically difficult to solve.

We therefore focused on the non-relativistic limit of the massless supersymmetric Abelian Higgs model, which involves multiple chiral fields with different gauge couplings. In this case the vacuum structure is richer, and non-trivial vacuum configurations satisfying $g_1|\varphi_1|^2+g_2|\varphi_2|^2=\xi$ minimize the energy of the system. Unlike the massive case, supersymmetry directly leads to first-order BPS equations obtained from the Killing spinor conditions. These equations admit vortex-type solutions, demonstrating that supersymmetry provides a natural mechanism for obtaining stable soliton configurations in the non-relativistic limit.

A central result of the massless model is that its vacuum moduli space has a weighted projective structure. By writing the gauge couplings as $g_a=g_0q_a$, with relatively prime positive integers $q_a$, the vacuum moduli space is described by the weighted projective space $W\mathbb{CP}^{1}_{(q_1,q_2)}$. This geometric structure plays an essential role in the classification of vortex solutions. At a generic point of the vacuum moduli space, where both scalar fields have nonzero asymptotic values, the two scalar phases must wind along the same broken $U(1)$ gauge orbit. This gives the compatibility condition $n_1/g_1=n_2/g_2$ and leads to the usual integer-winding BPS vortices with quantized magnetic flux.

The weighted projective structure also leads to fractional vortex configurations. For nontrivial weights, $W\mathbb{CP}^{1}_{(q_1,q_2)}$ contains orbifold fixed points at which one of the two scalar vacuum expectation values vanishes. At these special points, the residual discrete gauge symmetry modifies the usual quantization condition for the $U(1)$ winding number. Consequently, the allowed winding can become fractional, of the form $\nu=k/q_1$ or $\nu=k/q_2$, depending on which orbifold fixed point is approached at spatial infinity. Thus the fractional winding vortices found in this model arise naturally from the orbifold structure of the vacuum moduli space.

Assuming radial symmetry, we reduced the BPS equations to a system of coupled differential equations describing the radial profiles of the scalar and gauge fields. For generic asymptotic vacua, the analysis gives integer-winding topological vortices whose scalar profiles approach the vacuum exponentially and whose magnetic fields are localized near the vortex core. For asymptotic vacua located at the orbifold fixed points, the same BPS equations admit fractional winding configurations. The numerical analysis confirms the existence of smooth vortex profiles, localized magnetic fields, and localized energy densities, in agreement with the analytic asymptotic behavior.

We also clarified the interpretation of configurations in which one scalar field vanishes asymptotically. Such solutions should not be regarded as ordinary nontopological or Q-ball type configurations in the present model. Rather, when the asymptotic vacuum lies at an orbifold fixed point of $W\mathbb{CP}^{1}_{(q_1,q_2)}$, they are naturally interpreted as fractional vortices associated with the weighted projective vacuum moduli space. In contrast, configurations in which all scalar fields vanish asymptotically are incompatible with the finite-energy vacuum condition and therefore do not give admissible localized solitons in the present setting.

We have also emphasized the physical relevance of the model by discussing how the vacuum structure of the theory may be related to that of the Ginzburg--Landau theory of superconductivity. This comparison illustrates how phenomenological features familiar from superconducting systems can be viewed from a symmetry-based field-theoretic framework. It further highlights the usefulness of null reduction as a systematic tool for constructing non-relativistic supersymmetric theories while preserving the analytic advantages of their relativistic counterparts.

An interesting direction for future work would be to extend this analysis to models with a larger number of chiral multiplets. In such cases, the vacuum moduli spaces are expected to become higher-dimensional weighted projective spaces, and richer classes of integer and fractional vortex solutions may appear. It would also be worthwhile to investigate possible generalizations including Chern--Simons gauge interactions or non-Abelian gauge groups, where the interplay between non-relativistic supersymmetry, gauge dynamics, and vortex moduli spaces may lead to new types of BPS solitons.

\section*{\bf Acknowledgement}

We thank APCTP, Pohang, Korea for their hospitality during the TRP [APCTP-2026-T04] from which this work greatly benefited. This work was supported by the National Research Foundation
of Korea(NRF) grant with grant number NRF-2022R1F1A1073053 (Y.K.), RS-2019-NR040081, RS-2025-00553127, and RS-2026-25490199 (O.K.), RS-2025-25413714 and RS-2024-00351197 (H.S.).

\appendix
\section{Relevance of Our Theoretical Model}
In order to verify the relation of the theoretical models studied in this paper,  with the Ginzburg-Landau theory of superconductivity,  let us introduce the Gibbs free energy density of those models in the presence of an applied magnetic field ${\bf H}$, which is given by
\begin{align}\label{GFE}
{\cal G}={\cal H}-{\bf H}\cdot{\bf B}.
\end{align}
We note that in the case of the non-trivial extremum point obtained in section 3, since ${\bf B}$ is vanishing, the Gibbs free energy density is the same as the Hamiltonian density, whereas, for the trivial extremum point, the Gibbs free energy density is given by
\begin{align}\label{gibbsFE}
{\cal G}_0={\cal H}_0-H^2=\frac{\xi^2}{2}-\frac12H^2.
\end{align}

  In order to be stable vacuum states, the non-trivial extremum point must have lower Gibbs free energy than that of the trivial extremum point. For the massive Abelian Higgs model, let us first consider the $\xi>0$ case. If $~0\le\xi\le \frac{m^2}{g}$, then the only possible extremum point is the trivial solution, regardless of the value of the applied magnetic field ${\bf H}$. However, if $\xi>\frac{m^2}{g}$, then both the trivial solution and the non-trivial solutions are possible, where one of them is stable minimum vacuum state and the other one is an unstable point of maxima, depending on the value of the applied magnetic field. To figure out which one is the vacuum state, let us set $\xi=(1-\alpha)\frac{m^2}{g}$, for $\alpha<0$ , as we did in section 3. Then we can write the Gibbs free energy of the non-trivial solution \eqref{nontriv+} and that of trivial solution \eqref{gibbsFE} as
\begin{align}
	{\cal G}_{vac}^+=(1-2\alpha)\frac{m^4}{2g^2}, \qquad {\cal G}_{0}=\left(1-2\alpha+\alpha^2 \right)\frac{m^4}{2g^2}-\frac12H^2={\cal G}_{vac}^++\frac{\alpha^2m^4}{2g^2}-\frac12H^2, 
\end{align}
We note that the non-trivial solution is the stable vacuum state (${\cal G}_{vac}^+<{\cal G}_{0}$), only if 
\begin{align}\label{Hc}
	\frac{\alpha^2m^4}{2g^2}-\frac12H^2>0.
\end{align}
Therefore, with $\xi>\frac{m^2}{g}$ and the applied magnetic field $H<H_c$, where the critical magnetic field is 
\begin{align}
H_c=\frac{|\alpha| m^2}{g},
\end{align}
the non-trivial vacuum is the favourable lowest energy stable vacuum. On the other hand, with $\xi>\frac{m^2}{g}$ and the applied magnetic field $H>H_c$, both the trivial and non-trivial solutions are possible extremum points, but the trivial vacuum is the favourable lowest energy stable vacuum.
 
The fact that the non-trivial solution can exist only for $\xi>\frac{m^2}{g}$ (or $\alpha<0$), this massive model can be identified as the Ginzburg-Landau theory of superconductivity, where $\alpha$ corresponds to the parameter $a$ (the coefficient of the quadratic term in the Landau expansion of the scalar potential), and $g^2$ corresponds to the parameter $b$ (the coefficient of the quartic term in the expansion). More accurately, if $\alpha>0$ (or $0\le\xi\le \frac{m^2}{g}$ ), the energy is minimized by the trivial symmetric solution $\varphi_1=\varphi_2=0$, which corresponds to the normal non-superconducting state (no Cooper pairs), whereas for $\alpha<0$ (or $\xi> \frac{m^2}{g}$ ), the energy is minimized by the non-trivial vacuum solution $|\varphi_1|^2=\frac{\xi}g-\frac{m^2}{g^2},~ \varphi_2=0$, which means formation of Cooper pairs (electron Cooper pairs), with $U(1)$ charge $-g$ and such solution corresponds to the superconducting state, where $|\varphi_1|^2$ represent the number density of the superconducting electrons.

Now let us consider the massive model with $\xi<0$.  In this case, if $~-\frac{m^2}{g}\le\xi\le 0$, the only possible equilibrium state is the trivial vacuum state for any value of the applied magnetic field. On the other hand, if $\xi<-\frac{m^2}{g}$, both the trivial solution and non-trivial solution are possible extremum points, one of them being stable equilibrium and the other one is an unstable point of maxima, depending on the value of the applied magnetic field. To sort this out, now we set $\xi=-(1-\alpha)\frac{m^2}{g}$, for $(\alpha<0)$, and express the energies as 
\begin{align}
	{\cal G}_{vac}^-=(1-2\alpha)\frac{m^4}{2g^2},\qquad {\cal G}_{0}=\left(1-2\alpha+\alpha^2\right)\frac{m^4}{2g^2}-\frac12H^2={\cal G}_{vac}^-+\frac{\alpha^2m^4}{2g^2}-\frac12H^2.
\end{align}
This implies that, if the applied magnetic field $H<H_c=\frac{|\alpha| m^2}{g}$, the non-trivial solution is the favourable lowest energy stable vacuum state, whereas, if the applied magnetic field $H>H_c$, the trivial vacuum is the favourable lowest energy stable vacuum state. 
 
In this case as well, the massive model can be identified with the Ginzburg-Landau theory of superconductivity. When $\alpha>0$ (or $-\frac{m^2}{g}\le\xi\le0$ ), the energy is minimized by the trivial solution $\varphi_1=\varphi_2=0$, which corresponds to the normal non-superconducting state (no Cooper pairs), whereas when $\alpha<0$ (or $\xi< -\frac{m^2}{g}$ ), the energy is minimized by the non-trivial vacuum solution $\varphi_1=0,~|\varphi_2|^2=-\frac{\xi}{g}-\frac{m^2}{g^2}$, which means formation of another Cooper pairs (the positron Cooper pairs), with $U(1)$ charge $+g$, and this solution as well corresponds to the superconducting state, where $|\varphi_2|^2$ represent the number density of the superconducting positron.

Finally, let us consider the case of the massless model. In this case, since the energy of the non-trivial solutions is zero $({\cal G}_{vac}=0)$, these solutions are the favourable lowest energy stable vacuum states if the energy of the trivial solution is positive (${\cal G}_{0}>0$), which is true if  the applied magnetic field  $H<H_c=|\xi|$. If the applied magnetic field $H>H_c$, then the trivial solution is the favourable lowest energy stable vacuum.

In order to use the massless model as the Ginzburg-Landau theory of superconductivity, we need to fix the signatures of $g_1$ and $g_2$. If we choose $g_1$ and $g_2$ to be both positive, the model is the Ginzburg-Landau theory of multi-gap superconductivity, where $|\varphi_1|^2$ and $|\varphi_2|^2$, which for $\xi>0$ satisfy the equation of an ellipse \eqref{MLvac}, represent the number densities of two superconducting charge carriers, with $U(1)$ charges $-g_1$ and $-g_2$, respectively. The same can be said for $g_1$ and $g_2$ both negative with $\xi<0$. However, if $g_1$ and $g_2$ are opposite in sign, since $|\varphi_1|^2$ and $|\varphi_2|^2$ are solutions to the equation of a vertical or horizontal hyperbola, depending on the sign of $\xi$, the model will be the Ginzburg-Landau theory of multi-gap superconductivity,  with two oppositely charged superconducting charge carriers.    Recalling that, the massless model with opposite signs of $g_1$ and $g_2$ does not support vortex solutions, which means such model can serve as a model for type I superconductivity but not that of type II.  

\end{document}